\documentstyle[tighten,preprint,amsfonts,%
epsf,eqsecnum,aps,prd]{revtex}

\begin{document}
\draft

\hyphenation{
mani-fold
mani-folds
geo-metry
geo-met-ric
iso-metry
iso-metric
iso-metries
}



\def\BbbR{{\Bbb R}}
\def\BbbZ{{\Bbb Z}}
\def\BbbC{{\Bbb C}}

\def\Rmatrix{{\sf R}}

\def\half{{\frac{1}{2}}}
\def\casehalf{{\case{1}{2}}}

\def\Gammared{\Gamma_{\rm red}}
\def\Gammaredplus{\Gamma_{\rm red}^+}

\def\Othree{{\rm O}(3)}
\def\SOthree{{\rm SO}(3)}

\def\Ofourone{{\rm O}(4,1)}
\def\SOfourone{{\rm SO}(4,1)}
\def\ISOtwoone{{\rm ISO}(2,1)}

\def\Ofour{{\rm O}(4)}
\def\SOfour{{\rm SO}(4)}

\def\Otwoone{{\rm O}(2,1)}
\def\Octwoone{{\rm O}_{\rm c}(2,1)}
\def\tOctwoone{{\widetilde{{\rm O}_{\rm c}}}(2,1)}

\def\otwoone{{\frak o}(2,1)}
\def\sltwor{{\frak sl}(2,\BbbR)}

\def\SLtwor{{\rm SL}(2,\BbbR)}

\def\IOtwoone{{\rm IO}(2,1)}
\def\IOctwoone{{\rm IO}_{\rm c}(2,1)}

\def\arcsinh{\mathop{\rm arcsinh}\nolimits}
\def\arccosh{\mathop{\rm arccosh}\nolimits}
\def\arctanh{\mathop{\rm arctanh}\nolimits}


\preprint{\vbox{\baselineskip=12pt
\rightline{AEI 1999-015}
\rightline{MZ-TH/99-31}
\rightline{gr-qc/9908025}}}
\title{(2+1)-dimensional 
Einstein-Kepler problem 
\\
in the 
centre-of-mass frame}
\author{Jorma Louko\footnote{%
Electronic address:
jorma.louko@nottingham.ac.uk. 
}}
\address{
Max-Planck-Institut f\"ur Gravitations\-physik,
Am M\"uhlenberg~5, 
D-14476 Golm, 
Germany
\\
and
\\
School of Mathematical Sciences, 
University of Nottingham, 
\\
Nottingham NG7 2RD, 
United Kingdom\footnote{%
Present address}
}
\author{Hans-J\"urgen Matschull\footnote{Electronic address:
matschul@thep.physik.uni-mainz.de}}
\address{
Institut f\"ur Physik, 
Johannes-Gutenberg-Universit\"at, 
\\
Staudingerweg~7, 
D-55099 Mainz, 
Germany 
}
\date{August 1999; revised January 2000}
\maketitle
\begin{abstract}%
We formulate and analyse the 
Hamiltonian dynamics of a pair of massive
spinless point particles in $(2+1)$-dimensional Einstein gravity by
anchoring the system to 
a conical infinity, isometric to the
infinity generated by a single massive but possibly spinning
particle. The
reduced phase space $\Gammared$ has dimension four and topology
$\BbbR^3 \times S^1$. $\Gammared$~is analogous to the phase space of a 
Newtonian two-body system in the centre-of-mass frame, and we find on
$\Gammared$ a canonical chart that makes this analogue explicit and
reduces to the Newtonian chart in the appropriate limit. 
Prospects for 
quantisation are commented on. 
\end{abstract}

\narrowtext

\section{Introduction}
\label{sec:intro}

Einstein gravity in 2+1 spacetime dimensions provides an arena in
which many of the conceptual features of $(3+1)$-dimensional Einstein
gravity appear in a technically simplified setting 
\cite{carlip-book}.
One of these simplifications is that in 2+1 dimensions the theory can
be consistently coupled to point particles. The spacetimes containing
point particles can be described in terms of $\IOtwoone$ holonomies
around nontrivial loops
\cite{deser1,deser2}, 
and there exists considerable work on the
global structure of these spacetimes
\cite{thooft-cmp88,gott1,ori,deser-jack-th-prl,%
carroll1,carroll2,cutler,%
thooft-closed1,thooft-closed2,head-gott}, 
much of it motivated by the
observation that the spacetimes may contain closed causal curves.
Several variational formulations of the dynamics have been
introduced, both 
for 
examining the classical solution space in its own right 
and also 
as a starting point for
quantisation 
\cite{carlip-scat,%
cap-cia-val-plb,cap-cia-val-npb,%
thooft-quantum,%
bel-cia-val-plb,bel-cia-val-npb95,bel-cia-val-npb96,%
welling-noncomm,welling-winding,%
matsch-well,meno-sem}. 

In a spatially open universe, a variational formulation of
Einstein gravity 
must specify boundary conditions at the infinity. 
In 3+1 dimensions, the spacelike infinity of an isolated system 
can be taken
asymptotically Minkowski, 
and one can introduce in the Hamiltonian formulation a falloff that
anchors the system to an 
asymptotic
Minkowski spacetime \cite{MTW-mass-angmom,regge-teitel,beig-om}. 
The four-momentum and angular
momentum of the system, defined as surface integrals at the infinity,
can be interpreted respectively as a constant timelike vector and a
constant spacelike vector in the asymptotic Minkowski spacetime, 
and the asymptotic Poincar\'e isometry group can be used to 
choose an asymptotic centre-of-mass Lorentz frame  
\cite{MTW-mass-angmom}. 

By contrast, 
in 2+1 dimensions the spacelike infinity of an isolated
system is not asymptotically Minkowski but conical
\cite{henneaux-conical,ash-vara}. 
The neighbourhood of the infinity has  
only two independent 
globally-defined Killing vectors, a timelike
one generating time translations and a spacelike one 
generating rotations, but none that could be understood as 
generating boosts or spatial translations. It follows that the
neighbourhood of the spacelike infinity contains 
information that 
defines an analogue of a centre-of-mass frame
also in 
2+1 dimensions: 
in the special case of a spacetime 
containing a single massive 
and possibly spinning point particle, the metric near the
infinity uniquely determines the locus of the particle world line
\cite{deser1}. 
However, as the $\IOtwoone$ holonomy around the infinity is
nontrivial, with a nontrivial $\Otwoone$ part, the `momentum' and
`angular momentum' cannot be understood as constant vectors in an
asymptotic $(2+1)$-dimensional Minkowski spacetime, and the analogue
of the centre-of-mass frame cannot be realised as a Lorentz
frame in an asymptotic Minkowski spacetime.

With point particle sources in a spatially open $(2+1)$-dimensional 
spacetime, there is thus a certain 
tension between two different viewpoints on the dynamics. On the one
hand, one expects the conditions at the conical 
infinity to be crucial for
defining what evolution means, for finding a Hamiltonian that 
generates this evolution, and for discussing symmetries 
and conservation laws. On the other hand, the relative motion of the
particles takes a simple form when expressed in small patches of
Minkowski geometry valid `between' the particle world lines, but not
valid globally, and in particular not valid 
in a neighbourhood of the infinity: 
this suggests formulating the dynamics first in terms of fields and
variables that are defined locally, 
in small patches valid between the particles, 
and only later glueing the patches into a more
globally-defined formulation that incorporates the 
conditions at the infinity. 
Each of the variational formulations of 
\cite{carlip-scat,%
cap-cia-val-plb,cap-cia-val-npb,%
thooft-quantum,%
bel-cia-val-plb,bel-cia-val-npb95,bel-cia-val-npb96,%
welling-noncomm,welling-winding,%
matsch-well,meno-sem} 
strikes a
different balance between these two viewpoints. An example near one
extreme is \cite{matsch-well}, which specifies the trajectory of a
single particle in terms of a reference point
in the spacetime and a reference frame at this point.  
The purpose of the present paper is to approach
the opposite extreme: we anchor the particle trajectories 
to the conical infinity at the very outset.

We concentrate on the case of two massive spinless particles, which
can be regarded as 
the Kepler problem in $(2+1)$-dimensional Einstein
gravity. Briefly put, we shall formulate and solve the Hamiltonian
dynamics of the $(2+1)$-dimensional Einstein-Kepler problem in 
[the $(2+1)$-dimensional analogue of] the centre-of-mass frame.

To state the technical 
input more precisely, we assume the two-particle
spacetime to have a spacelike infinity whose neighbourhood is
isometric to a neighbourhood of the infinity in the spacetime of a
single massive but possibly spinning 
point particle \cite{deser1}, with a
defect angle less than~$2\pi$. This is equivalent to assuming that the
relative velocity of the two particles is less than the critical
velocity found in \cite{gott1,ori,deser-jack-th-prl}, and it implies that
the spacetime has no closed causal curves. 
The neighbourhood of the infinity admits a coordinate system that is
adapted to the isometries, and these conical coordinates foliate the
neighbourhood of the infinity by spacelike surfaces. We adopt these
conical coordinates as the asymptotic structure to which the
Hamiltonian dynamics will be anchored.

We take advantage of the well-known description of 
the two-particle spacetimes in
terms of a piece of Minkowski geometry between the particle world lines
\cite{gott1,ori,deser-jack-th-prl,carroll1,carroll2,cutler}. 
We first translate this description into one that 
relates the
world lines of the particles to the conical infinity.  We then use 
this explicit form of the classical solutions to reduce the
first-order gravitational action, 
and we find on the reduced phase space a canonical chart analogous 
to that in a nongravitating two-particle
system in the centre-of-mass frame. As expected from 
the nongravitating case, the reduced phase space has dimension four. 
The reduced Hamiltonian is bounded both above and
below, in agreement with the general arguments of 
\cite{henneaux-conical,ash-vara}, 
it has the correct relativistic 
test particle limit when the mass 
of one particle is small, and it has the correct
nonrelativistic limit when the masses and velocities of both particles 
are small. The
functional form of the Hamiltonian is nevertheless complicated, given 
only implicitly through the solution to a certain transcendental
equation. 
Quantising this reduced Hamiltonian theory would thus seem to present a
substantial challenge. 

The plan of the paper is as follows. In section
\ref{sec:spinning-point} we review the spacetime of a single 
spinning point
particle and introduce the conical coordinates. Section 
\ref{sec:spacetimes} describes the two-particle spacetimes as anchored 
to the conical infinity. Section \ref{sec:action-connection} recalls 
the first-order action 
\cite{matsch-well}
of $(2+1)$-dimensional 
Einstein gravity coupled to massive
point particles. The action is reduced in 
section~\ref{sec:reduction}, and the reduced phase space is analysed
in section~\ref{sec:com-coords}. Section \ref{sec:discussion} contains
brief concluding remarks. 

We use units in which $c = 4\pi G = 1$ (Planck's constant will not
appear). 
The Hamiltonian of a conical spacetime is in these units equal to half 
of the defect angle \cite{matsch-well}, 
and the mass of a point particle is then by definition 
half the defect
angle at the particle world line.  
$\Octwoone$ and $\IOctwoone$ stand
respectively for the connected components of $\Otwoone$ and
$\IOtwoone$.

\section{Spacetime of a single spinning point particle}
\label{sec:spinning-point}

In this section we briefly review 
the spacetime of a single spinning point
particle \cite{deser1}.  The main purpose
is to establish the notation, in particular the conical
coordinates~(\ref{conical-metric}). 

Let $M$ be the $(2+1)$-dimensional spacetime obtained by removing a
timelike geodesic from $(2+1)$ Minkowski space, 
and let ${\tilde M}$ be the
universal covering space of~$M$. We introduce on ${\tilde M}$ the
global coordinates $(T,R,\theta)$ in which the metric reads 
\begin{equation}
ds^2 = - dT^2 + dR^2 + R^2 d\theta^2
\ \ , 
\label{metric-TRtheta}
\end{equation}
such that $R>0$, $-\infty<T<\infty$, and $-\infty<\theta<\infty$. 
The only linearly-independent globally-defined Killing vectors on
${\tilde M}$ are 
$\partial_T$ and~$\partial_\theta$. 


Consider on ${\tilde M}$ the isometry 
$J := \exp 
\left( 
2\pi S \partial_T
+ 
2\pi \alpha \partial_\theta
\right)$, 
\begin{equation}
J : 
(T,R,\theta) \mapsto (T + 2\pi S,R,\theta+2\pi \alpha)
\ \ , 
\label{J-action}
\end{equation}
where $\alpha>0$ and $S\in\BbbR$. 
We interpret ${\tilde M}/J$ as the spacetime
generated by a spinning point particle at $R=0$ \cite{deser1}.
The mass of the particle in our units is $\pi(1-\alpha)$, 
and we refer to 
$S$ as the spin.
We take the mass
to be positive, and we thus have $0<\alpha<1$.

${\tilde M}/J$~can be described in terms of a fundamental domain
and an identification across its boundaries. 
As $\alpha<1$, this
fundamental domain can be embedded in~$M$, and the 
identification is then a $(2+1)$-dimensional Poincar\'e
transformation, consisting of a 
$2\pi\alpha$ rotation about the removed timelike
geodesic and a $2\pi S$ translation in the direction
of this geodesic. One choice for the
fundamental domain is the wedge $0<\theta<2\pi\alpha$. 

We refer to $2\pi\alpha$ as the opening angle and to $\delta := 
2\pi(1-\alpha)$
as the defect angle. When $S=0$, ${\tilde
  M}/J$ is the product spacetime of the time dimension and a
two-dimensional cone, and this terminology 
conforms to the standard terminology
for two-dimensional conical geometry.

We introduce on ${\tilde M}$ the coordinates $(t,R,\varphi)$ by 
\begin{mathletters}
\label{transf}
\begin{eqnarray}
&&T = t + S\varphi
\ \ ,
\\
&&\theta = \alpha \varphi
\ \ , 
\end{eqnarray}
\end{mathletters}
so that $R>0$, $-\infty<T<\infty$, and $-\infty<\theta<\infty$, and
the 
metric reads 
\begin{equation}
ds^2 = - (dt + S d\varphi)^2 + dR^2 + \alpha^2 R^2 d\varphi^2
\ \ . 
\label{conical-metric}
\end{equation}
In these coordinates 
$J = \exp (2\pi\partial_\varphi)$, so that 
\begin{equation}
J: (t,R,\varphi) \mapsto (t,R,\varphi+2\pi)
\ \ .
\label{corot-iddent}
\end{equation}
With the usual abuse of terminology, 
we refer to
$(t,R,\varphi)$ with the identification 
$(t,R,\varphi)\sim(t,R,\varphi+2\pi)$ as the 
{\em conical coordinates\/}
on~${\tilde M}/J$. The only linearly-independent Killing vectors on 
${\tilde M}/J$ are 
$\partial_t = \partial_T$ and $\partial_\varphi
= \alpha\partial_\theta + S \partial_T$. 

The conical coordinates on ${\tilde M}/J$ are uniquely defined 
up to the isometries generated by $\partial_t$
and~$\partial_\varphi$. $\partial_t$~is 
timelike, while 
$\partial_\varphi$ is spacelike for
$R>|S|/\alpha$, null for $R=|S|/\alpha$, and timelike for
$R<|S|/\alpha$. 
One can
think of $\partial_t$ as the generator of time translations
and $\partial_\varphi$ as the generator of rotations. 
${\tilde M}/J$~has closed causal curves, 
but none of them are 
contained in the region $R>|S|/\alpha$.

\section{Two-particle spacetimes with a conical infinity}
\label{sec:spacetimes}

In this section we describe the $(2+1)$-dimensional Einstein
spacetimes with a pair of 
massive spinless point particles, assuming the spacelike infinity to
be 
isometric to that of a single spinning point particle spacetime. 
In subsection \ref{subsec:local-descr} we recall the well-known 
description
in terms of a patch 
of Minkowski geometry between the particle world lines 
\cite{gott1,ori,deser-jack-th-prl,carroll1,carroll2,cutler}.
In the remaining three subsections we 
translate this description into one in which the particle world lines
are specified with respect to the conical infinity.

\subsection{Local description}
\label{subsec:local-descr}

We label the particles by the index $i=1,2$.
If the spacetime contains a collision of the particles, we consider 
either 
the part outside the 
causal past of the collision or the part that is 
outside the causal future of
the collision. 

In a neighbourhood of the world line of each particle, 
the geometry is the
spinless special case of the conical geometry of
section~\ref{sec:spinning-point}. 
The defect angles at the particles are $\delta_i :=
2\pi(1-\alpha_i)$. 
We regard these defect angles as prescribed
parameters, and we write $c_i := \cos(\delta_i/2)$, 
$s_i := \sin(\delta_i/2)$. We take the masses of the particles to be
positive, which implies $0<\delta_i<2\pi$, 
and the condition that
the far-region geometry be conical implies 
$\delta_1 + \delta_2 < 2\pi$ \cite{deser-jack-th-prl}.  
It follows that $s_i>0$ and $c_1+c_2>0$. 

The description of the spacetime in terms of a sufficiently small
piece $\tilde{D}$ of Minkowski spacetime 
`between' the particles is well known
\cite{gott1,ori,deser-jack-th-prl,carroll1,carroll2,cutler}.  
Each particle world line is a timelike
geodesic on the boundary of~$\tilde{D}$. 
The boundary segments of $\tilde{D}$ can in most cases be chosen 
as timelike plane segments joining at the particle world lines; the
only exception is when 
the particle world lines are not in the same timelike plane 
in $\tilde{D}$ and one 
of the defect angles is greater than~$\pi$, in which case the boundary 
segments need to be chosen suitably twisted 
(see subsection \ref{subsec:spinning} below). 
The 
identification across the two boundary segments joining at the world
line of particle $i$ is a rotation about this world line by the
angle~$\delta_i$. 
The effect of encircling in the spacetime {\em both\/} particle world
lines appears in the Minkowski geometry of $\tilde{D}$
as the composition of the two rotations, and this composition is a
Poincar\'e transformation that may in 
general have a translational component as well as a
Lorentz-component. Assuming the far-region geometry to 
be conical implies that the
Lorentz-component is a rotation (and
not a null rotation or a boost); 
this condition is equivalent to 
\cite{deser-jack-th-prl} 
\begin{equation}
-1 < c_1 c_2 - s_1 s_2 \cosh\beta
\ \ , 
\label{beta-ineq}
\end{equation}
where $\beta$ denotes the relative boost parameter of 
the particle trajectories in~$\tilde{D}$. 

Our task is to translate this description into one 
anchored to the conical infinity. We proceed in the following 
steps: 

$\bullet$
Cut $\tilde{D}$ into two along a suitably-chosen timelike surface
connecting the particle world lines; 

$\bullet$
Rotate the two halves of $\tilde{D}$ about the world line of particle 
$1$ so that the wedge originally at particle $1$ closes; 

$\bullet$
In the resulting new fundamental domain, find a set of 
Minkowski coordinates $(T,X,Y)$ and cylindrical
Minkowski coordinates $(T,R,\theta)$
in which the identification of the infinity-reaching 
boundary segments 
has the form~(\ref{J-action}); 

$\bullet$
Introduce the conical coordinates 
$(t,R,\varphi)$ (\ref{transf})  
in a neighbourhood of the
infinity, extend this neighbourhood inward 
to the particle trajectories, and
read off the particle trajectories in the conical
coordinates. 

When the particle trajectories are in the same timelike plane 
in the Minkowski geometry of~$\tilde{D}$, the boundary segments of
$\tilde{D}$ and the timelike surface along which $\tilde{D}$ is cut
can each be chosen to be in a timelike plane, 
and implementing the above steps is relatively straightforward. 
However, when the
particle trajectories in the Minkowski geometry of $\tilde{D}$ are not
in the same timelike plane, the timelike surface along which
$\tilde{D}$ is cut cannot be chosen to be in a plane, and
when in addition one defect angle is larger than~$\pi$, the
boundary segments of $\tilde{D}$ cannot be chosen to be in timelike
planes arbitrarily far into the future and the past 
(see subsection \ref{subsec:spinning} below). 
The geometry in this latter case is thus quite subtle. 

We divide the analysis into three qualitatively different cases,
proceeding from the simplest one to the most intricate one. 
First, when the particle
trajectories in the Minkowski geometry of $\tilde{D}$ are parallel,
the spacetime is static and has in particular vanishing spin. Second,
when the particle trajectories in the Minkowski geometry of
$\tilde{D}$ are in the same timelike plane but not parallel, the
spacetime has a vanishing spin but is not static, and it contains a
collision of the particles. Third, when the trajectories in the
Minkowski geometry of $\tilde{D}$ are not in the same timelike plane,
the spacetime has a nonvanishing spin: this is the `generic' case.
We devote a separate subsection to each case. 

We record here the result, noted in 
\cite{deser-jack-th-prl} and to be verified below in all our
three cases, that the defect angle $\delta$ of the 
far-region conical geometry is the unique solution 
in the interval $0<\delta<2\pi$ to the equation 
\begin{equation}
\cos(\delta/2) = c_1 c_2 - s_1 s_2 \cosh\beta
\ \ .
\label{delta-deter}
\end{equation}
Note that this implies $\delta\ge \delta_1 + \delta_2$, 
equality holding iff $\beta=0$. 
The parameter $\alpha$ of the far-region conical
geometry is 
\begin{equation}
\alpha = 1 - \delta/(2\pi)
\ \ .
\label{alphadelta-note}
\end{equation}

\subsection{$\beta=0$: Static spacetimes}
\label{subsec:static}

When $\beta=0$, the particle world lines 
on the boundary of $\tilde{D}$
are parallel in the Minkowski geometry of~$\tilde{D}$. 
We introduce in $\tilde{D}$ Minkowski coordinates 
$({\tilde{t}},{\tilde{x}},{\tilde{y}})$ 
such that $\partial_{\tilde{t}}$ points to the
future, the world line of particle $1$ is
$({\tilde{t}},{\tilde{x}},{\tilde{y}}) = (\lambda,0,0)$, and the 
world line of particle $2$ is
$({\tilde{t}},{\tilde{x}},{\tilde{y}}) = 
(\lambda, -a,0)$, where $a$ is a positive constant. 
Here $\lambda$ stands for a proper time parameter individually on each
world line. 

\begin{figure}
\begin{center}
\vglue 3 cm
\leavevmode
\epsfysize=10cm
\epsfbox{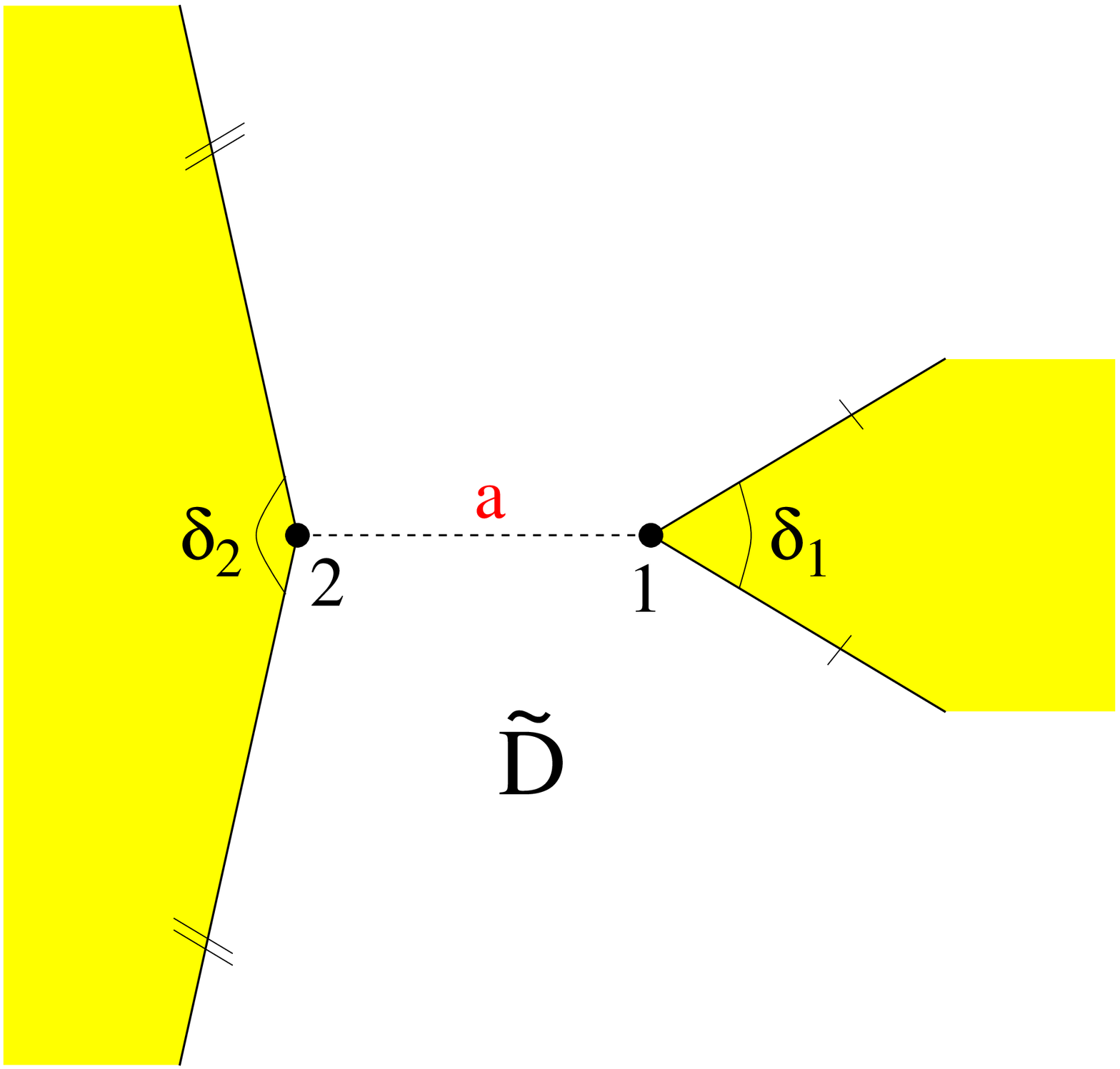}
\end{center}
\vskip 3 cm
\caption{%
A~constant ${\tilde{t}}$ surface of the fundamental domain 
for a static spacetime. The parts not belonging to the fundamental
domain are shown as shaded. 
The particle world lines are 
orthogonal to the plane of the paper, and the orthogonal spatial
distance of the world lines is~$a$. 
The boundary segments marked by a single (double) stroke 
are identified by a
rotation about particle $1$ ($2$) 
by the angle $\delta_1$
($\delta_2$).}
\label{fig:staticwedges}
\end{figure}

We choose $\tilde{D}$ so that its boundary segments are half-planes
bounded by the particle world lines and its intersection with the
surface ${\tilde{t}}=0$ is as shown in
figure~\ref{fig:staticwedges}. 
The spacetime is clearly static. 

To obtain 
an equivalent fundamental domain that is better adapted to the
infinity, we proceed as outlined in
subsection~\ref{subsec:local-descr}. 
We first work in the plane
of figure \ref{fig:staticwedges} and then extend into the time
dimension by staticity. 

\begin{figure}
\begin{center}
\vglue 3 cm
\leavevmode
\epsfysize=10cm
\epsfbox{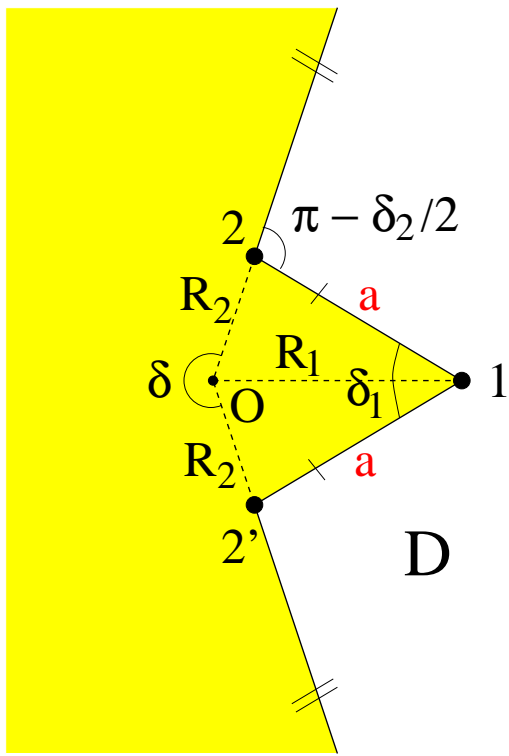}
\end{center}
\vskip 3 cm
\caption{%
A~constant $T$ surface of the fundamental domain $D$
for the static spacetime of figure~\ref{fig:staticwedges}. The
corner labelled $1$ is at the first particle, and the corners
labelled $2$ and $2'$ are at the second particle. 
The particle world lines are orthogonal to the surface.
The single-stroked boundaries 
are identified by a
rotation about particle $1$ 
by the angle~$\delta_1$, and the double-stroked 
boundaries
are identified by a
rotation about the (fictitious) origin $O$ 
by the angle $\delta = \delta_1+\delta_2$. 
Note that $O$ lies outside~$D$. 
The (fictitious) distances $R_i$ of the
particles from $O$ are given by~(\ref{Rvalues-static}).}
\label{fig:staticfundomain}
\end{figure}

We cut
the planar fundamental domain of 
figure \ref{fig:staticwedges} into two 
along the straight line connecting the particles, 
and we rotate the two 
halves with respect to each other about particle $1$ 
so that the wedge on the right
closes. The
new planar fundamental domain is shown in 
figure~\ref{fig:staticfundomain}:
the corner labelled $1$ is at the first particle, while the
corners labelled $2$ and $2'$ are both at the second particle. We
introduce in this new planar fundamental domain 
polar coordinates
$(R,\theta)$ in which the boundary segments
from $2$ and $2'$ to infinity are at
constant~$\theta$: the origin of these polar coordinates lies outside
the domain and is labelled in 
figure \ref{fig:staticfundomain}
by~$O$. 
We denote the value of $R$ at corner $1$ by~$R_1$, and that 
at corners $2$ and $2'$ by~$R_2$. Elementary planar geometry yields 
\begin{mathletters}
\label{Rvalues-static}
\begin{eqnarray}
&&R_1 =
\frac{a s_2}{\sin(\delta/2)}
\ \ ,
\\
&&R_2 =
\frac{a s_1}{\sin(\delta/2)}
\ \ ,
\end{eqnarray}
\end{mathletters}
where $\delta = \delta_1 + \delta_2$. 
Choosing the corner $1$ to be at $\theta=0$, the corners $2$
and $2'$ are respectively at $\theta = \pm (\pi - \delta/2)$. 

\begin{figure}
\begin{center}
\vglue 3 cm
\leavevmode
\epsfysize=10cm
\epsfbox{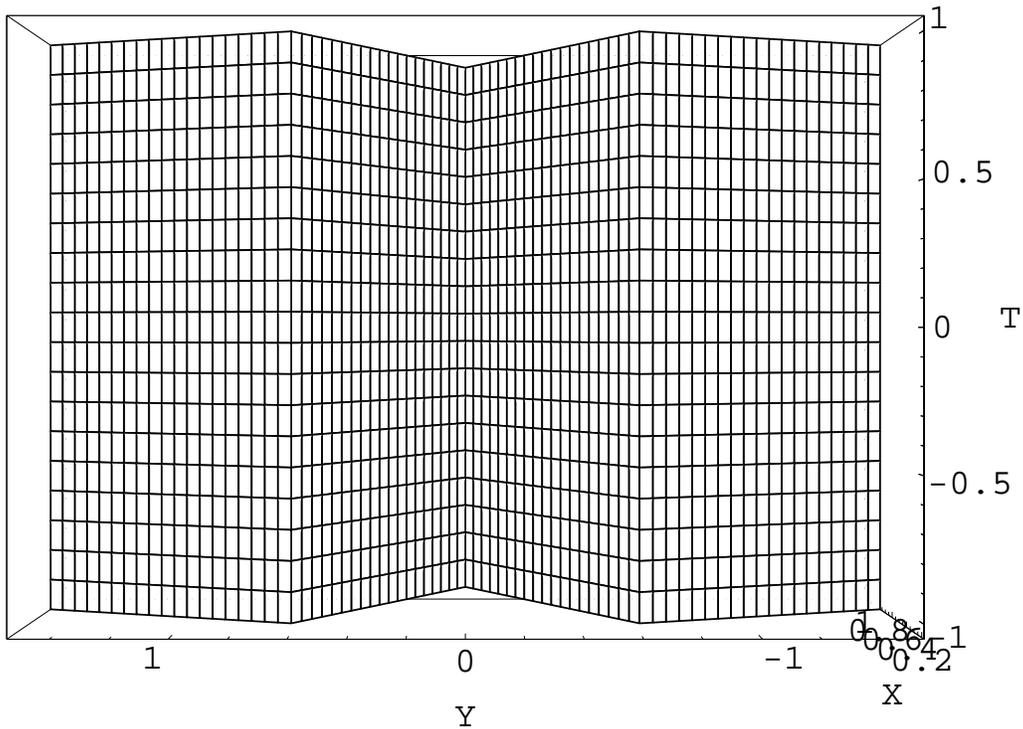}
\end{center}
\vskip 3 cm
\caption{%
The 
boundary of the fundamental domain $D$
for the static spacetime in the Minkowski coordinates $(T,X,Y)$. 
$D$~is behind the boundary, and the parameters are 
$\delta_1 = 2\pi/5$, $\delta_2 = 4\pi/5$, and $a=1$. 
The viewpoint is on the negative $X$-axis. 
The grid on the two segments between the particles is adapted to the
identification of these segments, and similarly for the grid on the
two segments reaching the infinity. Note that the identifications
do not affect the Minkowski time coordinate~$T$. 
}
\label{fig:static3d-fundomain}
\end{figure}

The new spacetime fundamental domain $D$ is the product of
the planar fundamental domain of 
figure \ref{fig:staticfundomain} and the time axis. 
In the cylindrical Minkowski coordinates 
$(T,R,\theta)$ (\ref{metric-TRtheta}) in~$D$, 
the two infinity-reaching boundary
segments of $D$ are 
in the half-planes $\theta = \pm (\pi
- \delta/2)$, 
and their identification is a pure rotation about the
(fictitious) axis at $R=0$. 
A~spacetime picture of $D$ is shown in
figure~\ref{fig:static3d-fundomain}. 
The spacetime near the infinity
is thus conical with vanishing spin, and $\delta$ is 
the defect angle. 
Note that this discussion verifies equation 
(\ref{delta-deter}) in the special case
$\beta=0$. 

We now introduce near the infinity the conical coordinates
$(t,R,\varphi)$ as in~(\ref{transf}), 
with $S=0$ and with $\alpha$ given by~(\ref{alphadelta-note}), 
except in that we add to 
$\varphi$ a constant $-\varphi_0$ that is defined modulo~$2\pi$. 
The conical coordinates are
valid in a neighbourhood of the infinity, 
and we can extend this neighbourhood inwards 
(in a $\varphi$-dependent way if $\delta_1\ne\delta_2$) so
that the boundary of the extended neighbourhood contains the
trajectories of both particles. 
The world line of particle 1 is then at 
$(t,R,\varphi) = (\lambda,R_1,\varphi_0)$, and that of 
particle 2 is at $(t,R,\varphi) = (\lambda,R_2,\varphi_0+\pi)$. 

To summarise, we have introduced a neighbourhood of the infinity
covered by the conical coordinates $(t,R,\varphi)$ and expressed the
particle trajectories as lines on the boundary of this neighbourhood.
The values of the radial coordinate $R$ at the particles are given
by~(\ref{Rvalues-static}), and the values of the angular coordinate
$\varphi$ are respectively $\varphi_0$ and~$\varphi_0+\pi$: in this
sense, the particles are diametrically opposite each other.  The
parameter $\varphi_0$ specifies the orientation of the two-particle
system relative to the conical coordinates, and spacetimes
differing only in the value of $\varphi_0$ are clearly isometric.

\subsection{$\beta\ne0$, $S=0$: 
Spacetimes with colliding particles}
\label{subsec:colliding}

When $\beta\ne0$ but the particle trajectories 
in the Minkowski geometry of
$\tilde{D}$ are in the same timelike plane, the spacetime contains 
a collision of the particles. We consider either 
the part outside the 
causal past of the collision or the part that is 
outside the causal future of
the collision. 

As in subsection~\ref{subsec:static}, we 
introduce in $\tilde{D}$ Minkowski coordinates 
$({\tilde{t}},{\tilde{x}},{\tilde{y}})$ in which 
the world line of particle $1$ is
$({\tilde{t}},{\tilde{x}},{\tilde{y}}) = (\lambda,0,0)$. 
We now choose these coordinates so that the world line of particle 
$2$ is
$({\tilde{t}},{\tilde{x}},{\tilde{y}}) = 
(\lambda\cosh\beta, - \lambda\sinh\beta,0)$ at ${\tilde{x}}<0$. 
The particles collide at 
$({\tilde{t}},{\tilde{x}},{\tilde{y}}) = (0,0,0)$. 
For $\beta>0$ ($\beta<0$, respectively), the particles
are receding (approaching). 

\begin{figure}
\begin{center}
\vglue 2 cm
\leavevmode
\epsfysize=10cm
\epsfbox{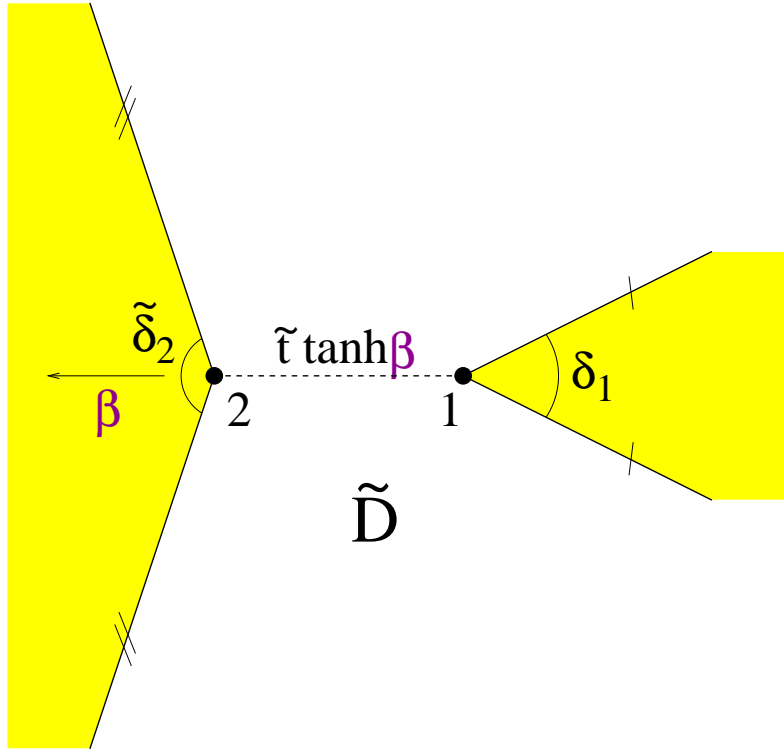}
\end{center}
\vskip 2 cm
\caption{%
A constant ${\tilde{t}}$ surface of the fundamental 
domain $\tilde{D}$ for a
spacetime with colliding particles. The world line of particle $1$ is
orthogonal to the plane of the paper, while the world line 
of particle $2$ has the boost parameter $\beta\ne0$.
For $\beta>0$ ($\beta<0$), particle $2$ is moving to
the left (right), and the surface shown is at ${\tilde{t}}>0$
(${\tilde{t}}<0$). The spatial distance of the particles in the
constant ${\tilde{t}}$ surface, along the dashed line, 
is ${\tilde{t}}\tanh\beta$. 
The angle ${\tilde{\delta}}_2$ is determined by 
$\delta_2$ and $\beta$ as 
the unique solution to
$\cos({\tilde{\delta}}_2/2) 
= c_2 \left(1 + s_2^2
  \sinh^2\beta\right)^{-1/2}$
in the interval 
$0<{\tilde{\delta}}_2<2\pi$. 
The single-stroked (double-stroked) boundaries 
are identified by a
rotation about the world line of particle $1$ ($2$) 
by the angle $\delta_1$~($\delta_2$). 
Note that equation 
(\ref{beta-ineq}) implies 
$\delta_1 + {\tilde{\delta}}_2<2\pi$, even
when one of the defect angles is greater than~$\pi$: 
this guarantees that the 
straight lines in the figure, 
at the boundaries of the~$\tilde{D}$, 
always reach the infinity without intersecting.
}
\label{fig:collwedges}
\end{figure}

We choose $\tilde{D}$ so that its intersection with the surface
${\tilde{t}}=0$ is as shown in figure~\ref{fig:collwedges}. The 
boundary segments of $\tilde{D}$ are continued to 
${\tilde{t}}\ne0$ in the half-planes
bounded by the respective particle world lines.

To find a fundamental domain better suited to the infinity, we
proceed as outlined in subsection~\ref{subsec:local-descr}. 
We cut $\tilde{D}$ 
into two along the totally geodesic timelike surface between 
the particle world lines, and we 
rotate the two halves 
with respect to each other about the world line
of particle $1$ so that the wedge originally at particle $1$
closes. In the resulting new fundamental domain~$D$, 
we introduce 
Minkowski coordinates $(T,X,Y)$ and cylindrical
Minkowski coordinates $(T,R,\theta)$
in which the identification of the infinity-reaching 
boundary segments 
has the form~(\ref{J-action}). 
We choose these coordinates so that particle
$1$ has always $\theta=0$ and the collision of the
particles is at $(T,X,Y)=(0,0,0)$. 

\begin{figure}
\begin{center}
\vglue 2 cm
\leavevmode
\epsfysize=10cm
\epsfbox{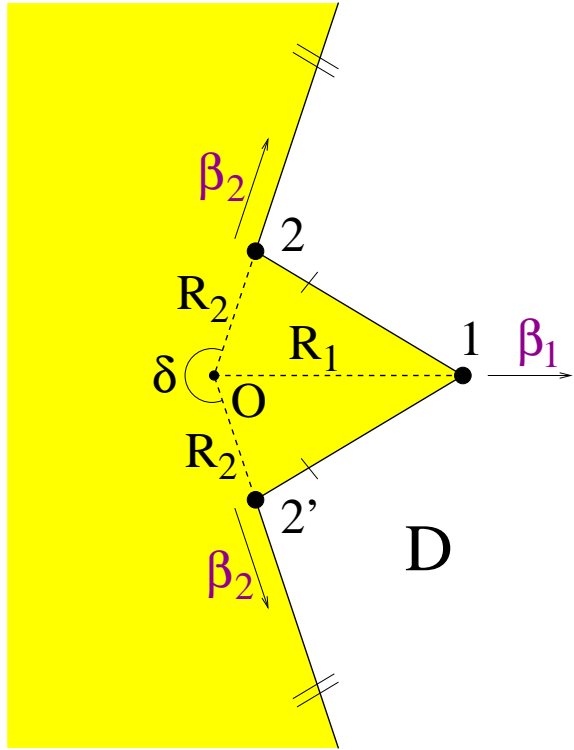}
\end{center}
\vskip 2 cm
\caption{%
A~constant $T$ surface of the fundamental domain $D$
for the spacetime with colliding particles. 
The corner labelled $1$ is at the first particle, and the corners
labelled $2$ and $2'$ are at the second particle. 
The world lines of the particles have the 
nonvanishing boost parameters $\beta_i$ 
(\ref{betaonetwo-def}) with respect to the Minkowski coordinates
$(T,X,Y)$. For $\beta>0$, $\beta_i$ and $T$ are positive and
the velocities are in the directions shown; for $\beta<0$, 
$\beta_i$ and $T$ are negative, and the directions are the
opposite. 
The single-stroked boundaries 
are identified by a 
rotation in the spacetime about the world line of particle $1$ 
by the angle~$\delta_1$. The double-stroked 
boundaries
are identified by~(\ref{J-action}). 
The distances of the
particles from $O$ in (the extension beyond $D$ of) 
the constant $T$ surface 
are $R_i := T \tanh\beta_i$.}
\label{fig:collfundomain}
\end{figure}

\begin{figure}
\begin{center}
\vglue 2 cm
\leavevmode
\epsfysize=12cm
\epsfbox{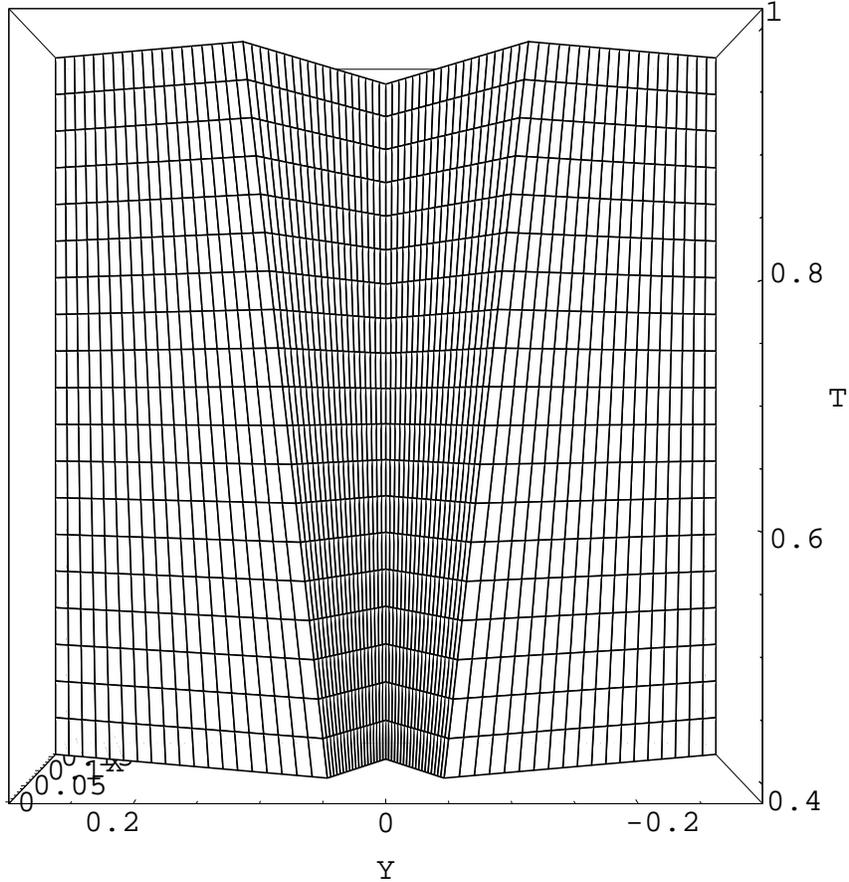}
\end{center}
\vskip 3 cm
\caption{%
The 
boundary of the fundamental domain $D$
for the colliding-particle spacetime 
in the Minkowski coordinates $(T,X,Y)$. 
The collision is in the past, 
$D$~is behind the boundary, 
and the parameters are 
$\delta_1 = 2\pi/5$, $\delta_2 = 4\pi/5$, and $\beta = 0.2$. 
The viewpoint is at $(T,X,Y) = (0.7,-2,0)$. 
The grid is adapted to the identifications of the boundaries 
as in figure~\ref{fig:static3d-fundomain}. 
Note that the identifications do not
affect the Minkowski time coordinate~$T$.
}
\label{fig:coll3d-fundomain}
\end{figure}

The algebra in finding $(T,R,\theta)$ is lengthy 
but straightforward. 
A~constant $T$ surface of $D$ is shown in
figure~\ref{fig:collfundomain}, with $T>0$ ($T<0$)
for $\beta>0$ ($\beta<0$).
The first particle is at the corner labelled~$1$, at 
$(T,R,\theta) = (T,T\tanh\beta_1,0)$, 
and the second particle is at the corners labelled 
$2$ and~$2'$, respectively at 
$(T,R,\theta) = \biglb(T,T\tanh\beta_2,\pm(\pi - \delta/2)\bigrb)$,
where $\delta$ is defined by (\ref{delta-deter}) and 
$\beta_i$ by 
\begin{mathletters}
\label{betaonetwo-def}
\begin{eqnarray}
&&\sinh\beta_1 =
\frac{s_2 \sinh\beta}{\sin(\delta/2)}
\ \ ,
\label{betaone-def}
\\
&&\sinh\beta_2 =
\frac{s_1 \sinh\beta}{\sin(\delta/2)}
\ \ . 
\end{eqnarray}
\end{mathletters}
$\beta_i$~are thus the boost parameters of the particle 
world lines with respect to the coordinates $(T,X,Y)$. 
The boundary of $D$ between edges $1$ and $2$ is
the timelike plane sector connecting these edges, and similarly
for the boundary between edges $1$ and~$2'$. 
The infinity-reaching boundaries are in the timelike planes 
$\theta=\pm(\pi - \delta/2)$, and their
identification is by the map (\ref{J-action}) with $S=0$. 
The spacetime near the infinity is thus conical with 
$S=0$, and the defect angle is~$\delta$. 
A~spacetime picture of $D$ 
(with $\beta>0$) is shown in 
figure~\ref{fig:coll3d-fundomain}. 

Near the infinity, we introduce conical coordinates
$(t,R,\varphi)$ as in~(\ref{transf}), except in that 
we replace $t$ by $t-t_0$ and 
$\varphi$ by $\varphi-\varphi_0$, 
where $t_0$ and $\varphi_0$ are constants, the
latter one defined modulo~$2\pi$. 
The resulting conical coordinates are
valid in a neighbourhood of infinity, and we extend this neighbourhood
inwards so
that the particle trajectories lie on its boundary. 
The trajectory of
particle $1$ is 
$(t,R,\varphi) =
(t_0 + \lambda\cosh\beta_1,\lambda\sinh\beta_1,\varphi_0)$, 
and that of particle
$2$ is $(t,R,\varphi) =
(t_0 + \lambda\cosh\beta_2,\lambda\sinh\beta_2,\varphi_0+\pi)$. 
The particles are thus again diametrically opposite 
at each~$t$, and the constant $\varphi_0$
is the conical angle of particle~$1$, specifying the orientation of
the two-particle system with respect to the infinity. The constant
$t_0$ is the value of the conical time at the collision of the
particles.

\subsection{$S\ne0$: Spinning spacetimes}
\label{subsec:spinning}

When the particle trajectories in the Minkowski geometry of
$\tilde{D}$ are not in the same timelike plane, we again 
introduce in $\tilde{D}$ Minkowski coordinates 
$({\tilde{t}},{\tilde{x}},{\tilde{y}})$ 
such that $\partial_{\tilde{t}}$ points to the
future and the world line of particle $1$ is
$({\tilde{t}},{\tilde{x}},{\tilde{y}}) = (\lambda,0,0)$. 
We now choose the coordinates so that 
the world line of particle $2$ is 
$({\tilde{t}},{\tilde{x}},{\tilde{y}}) =
(\lambda\cosh\beta,-a,-\lambda\sinh\beta)$, where $a>0$ and
$\beta\ne0$. 

\begin{figure}
\begin{center}
\vglue 2 cm
\leavevmode
\epsfysize=10cm
\epsfbox{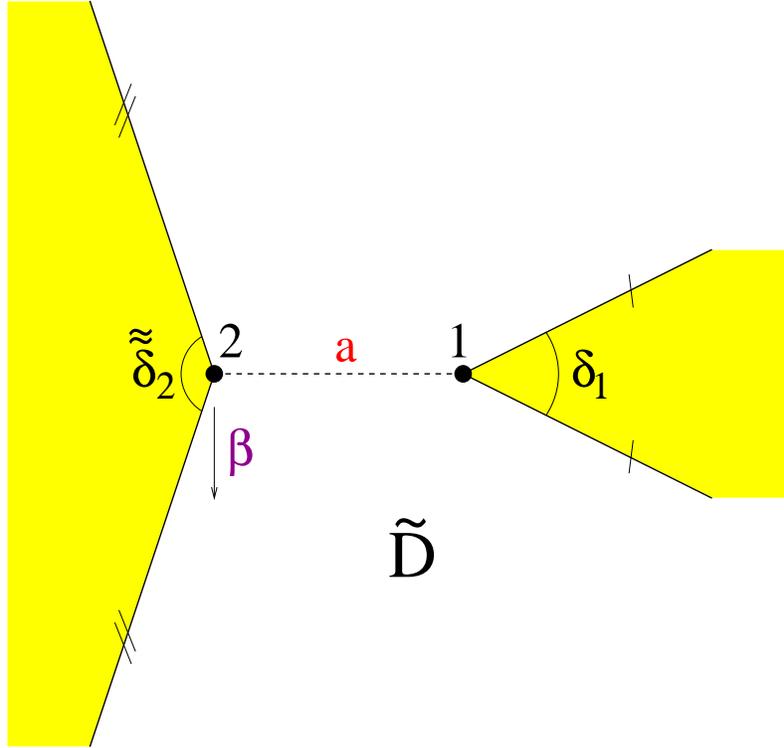}
\end{center}
\vskip 2 cm
\caption{%
The surface ${\tilde{t}}=0$ of the fundamental 
domain $\tilde{D}$ for a spinning spacetime. 
The angle ${\tilde{\tilde{\delta}}}_2$ 
is determined by $\delta_2$ and $\beta$ as
the unique solution to
$\cos({\tilde{\tilde{\delta}}}_2/2) = c_2 \left(1 - s_2^2
  \tanh^2\beta\right)^{-1/2}$
in the interval 
$0<{\tilde{\tilde{\delta}}}_2<2\pi$. 
The extension of $\tilde{D}$ beyond the surface ${\tilde{t}}=0$ is
described in the text. In the spacetime, single-stroked
(double-stroked) boundary segments are 
identified by a
rotation about the world line of particle $1$ ($2$) 
by the angle $\delta_1$~($\delta_2$). 
The rotation about particle $1$ 
takes the ${\tilde{t}}=0$ sections of the single-stroked boundary
segments to each other, but the rotation about particle $2$ does
not take the ${\tilde{t}}=0$ sections of the double-stroked boundary
segments to each other. 
}
\label{fig:spinwedges}
\end{figure}

We choose the intersection of $\tilde{D}$ with the surface
${\tilde{t}}=0$ as shown in figure~\ref{fig:spinwedges}. If neither
defect angle is greater than~$\pi$, one possible choice for the
boundary segments of $\tilde{D}$ would be 
to continue them off 
the ${\tilde{t}}=0$ surface as 
timelike half-planes bounded by the respective particle world
lines \cite{gott1,cutler}. 
If one defect angle is greater than~$\pi$, such 
half-planes would however 
eventually intersect the trajectory of
the other particle, and the boundary segments need 
to be chosen 
suitably twisted. 
It is fortunately not necessary to specify here precisely how the 
boundary segments of $\tilde{D}$ are chosen: 
we shall construct a new fundamental domain
$D$ as 
outlined in subsection~\ref{subsec:local-descr}, 
and we shall specify the boundary segments of $D$ in a way that is
more easily described directly in terms of~$D$. 
In particular, we choose the infinity-reaching boundary segments of
$D$ not to be in timelike planes for any values
of the defect angles. 

\begin{figure}
\begin{center}
\vglue 2 cm
\leavevmode
\epsfysize=10cm
\epsfbox{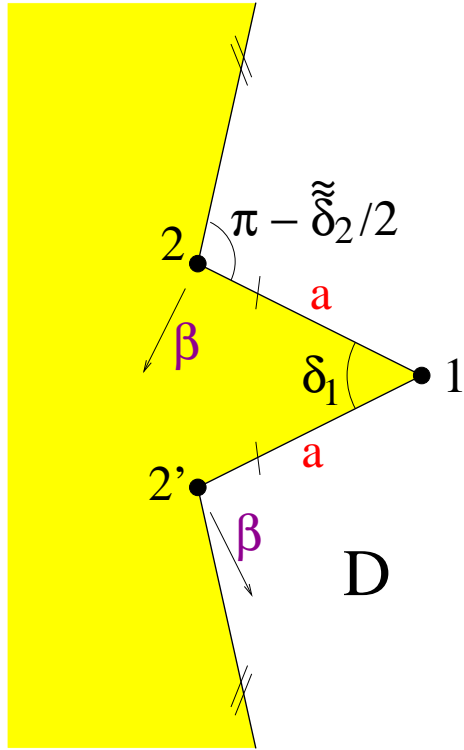}
\end{center}
\vskip 2 cm
\caption{%
A surface of the fundamental 
domain $D$ for a spinning spacetime.
We have first cut
$\tilde{D}$ into two along a timelike surface
that connects the world
lines of the particles, as explained in the text. 
We have then rotated the two halves with respect to
each other about the world line of particle $1$ so that the wedge
originally at particle~$1$, in figure \ref{fig:spinwedges}
on the right, closes. The corner labelled $1$ is at the first particle, 
and the corners labelled $2$ and $2'$ are at the second particle. 
In the spacetime, 
the single-stroked (double-stroked) boundary segments 
are identified as described in the text. 
}
\label{fig:spinwedges-rotated}
\end{figure}

We first choose in $\tilde{D}$ a timelike surface that
connects the world lines of the two particles. We take this
surface to contain the spacelike geodesic that connects the
particles at ${\tilde{t}}=0$, shown as the dashed line in 
figure~\ref{fig:spinwedges}; the choice of the surface for
${\tilde{t}}\ne0$ will be specified shortly. 
We cut $\tilde{D}$ into two along this surface, and we 
rotate the two
halves with respect to each other about the world line of particle $1$ 
so that the wedge originally at particle~$1$, 
in figure \ref{fig:spinwedges}
on the right, closes. 
The ${\tilde{t}}=0$ surface of the resulting domain $D$
is shown in figure~\ref{fig:spinwedges-rotated}. The corner 
labelled $1$ is at the first particle, 
and the corners labelled $2$ and $2'$ are both at the second
particle. 

Finding in $D$
Minkowski coordinates $(T,X,Y)$ and cylindrical
Minkowski coordinates $(T,R,\theta)$ 
in which the identification of the infinity-reaching 
boundary segments 
has the form~(\ref{J-action}) is again
lengthy but straightforward. 
Let $\delta$ be defined by~(\ref{delta-deter}), 
let $\beta_i$ be defined by~(\ref{betaonetwo-def}), 
and let 
\begin{mathletters}
\label{rhos-def}
\begin{eqnarray}
&&\rho_1 :=
\frac{a s_2 \cosh\beta_2}{\sin(\delta/2)}
\ \ ,
\\
&&\rho_2 :=
\frac{a s_1 \cosh\beta_1}{\sin(\delta/2)}
\ \ .
\end{eqnarray}
\end{mathletters}
Let $(T_1,R_1,\theta_1)$ stand for the cylindrical 
Minkowski coordinates of the 
edge labelled $1$ in
figure~\ref{fig:spinwedges-rotated}, and let 
$(T_2,R_2,\theta_2)$ and $(T_{2'},R_{2'},\theta_{2'})$ 
similarly stand for 
the respective cylindrical Minkowski 
coordinates of the
edges labelled $2$ and~$2'$. 
We choose the point ${\tilde{t}}=0$ at edge 
$1$ to be at $T_1=0=\theta_1$. 
We then find 
\begin{mathletters}
\label{Rs-of-sigma}
\begin{eqnarray}
R_1 &=& \frac{\rho_1}{\cos\sigma}
\ \ ,
\\
\noalign{\smallskip}
R_2 = R_{2'} &=& \frac{\rho_2}{\cos\sigma}
\ \ ,
\label{Rtwos-of-sigma}
\end{eqnarray}
\end{mathletters}
\begin{mathletters}
\label{Ts-of-sigma}
\begin{eqnarray}
&& T_1 = 
\frac{a \cosh \beta_1 \cosh \beta_2 }{\sinh\beta} \tan\sigma
\ \ ,
\\
&&
T_2 = 
a \left[
\frac{\cosh \beta_1 \cosh \beta_2 }{\sinh\beta} \tan\sigma
+ \frac{s_1 s_2 \sinh\beta}{\sin(\delta/2)}
\right]
\ \ ,
\\
&&
T_{2'} = 
a \left[
\frac{\cosh \beta_1 \cosh \beta_2 }{\sinh\beta} \tan\sigma
- \frac{s_1 s_2 \sinh\beta}{\sin(\delta/2)}
\right]
\ \ ,
\end{eqnarray}
\end{mathletters}
and 
\begin{mathletters}
\label{thetas-of-sigma}
\begin{eqnarray}
&& \theta_1 = \sigma
\ \ ,
\\
&&
\theta_2 = \sigma + \alpha \pi
\ \ ,
\\
&&
\theta_{2'} = \sigma - \alpha \pi
\ \ ,
\end{eqnarray}
\end{mathletters}
where $\sigma \in (-\casehalf \pi, \casehalf\pi)$ 
is a parameter along
each of the three edges. For 
$\beta>0$ ($\beta<0$), 
$\sigma$~grows towards the
future (past). 
The map that identifies the infinity-reaching 
boundary segments of~$D$, 
taking in particular edge $2'$ to
edge~$2$,
is (\ref{J-action}) with 
\begin{equation}
S := 
\frac{a s_1 s_2 \sinh\beta}{\pi \sin(\delta/2)}
\ \ . 
\label{S-def}
\end{equation}
In the Minkowski coordinates $(T,X,Y)$, 
the tangent vector of edge $1$ is
$(\cosh\beta_1, \sinh\beta_1,0)$, 
the tangent vector of edge $2$ is 
$\biglb(\cosh\beta_2, 
-\sinh\beta_2 \sin(\delta/2),-\sinh\beta_2 \cos(\delta/2)\bigrb)$, 
and the tangent vector of  edge $2'$ is 
$\biglb(\cosh\beta_2, \sinh\beta_2 \sin(\delta/2),-\sinh\beta_2
\cos(\delta/2)\bigrb)$. 
$\beta_i$~are thus the boost parameters of the
respective edges with respect to the coordinates $(T,X,Y)$, as in the
colliding-particle spacetimes of subsection~\ref{subsec:colliding}. 

Having found the edges on the boundary of~$D$, 
we are ready
to specify the rest of the boundary. First, note that the
identification of edges $2$ and $2'$ takes a point on edge $2'$ with a
given value of $\sigma$ to a point on edge $2$ with the {\em same\/}
value of~$\sigma$. 
We can therefore choose the boundary segment between edges 
$1$ and $2$ to consist of 
geodesics joining 
edge $1$ to edge $2$ at the {\em same\/} 
value of $\sigma$ on the two edges, and
similarly for the boundary segment between edges $1$ and~$2'$: 
this specifies the way how $\tilde{D}$ was cut in two. 
All these geodesics are spacelike. 
Finally, we take the boundary segment from edge $2$ to
infinity to consist of half-lines at constant $T$ and~$\theta$, and
similarly for the boundary segment from edge 
$2'$ to infinity. 

\begin{figure}
\begin{center}
\vglue 2 cm
\leavevmode
\epsfysize=11cm
\epsfbox{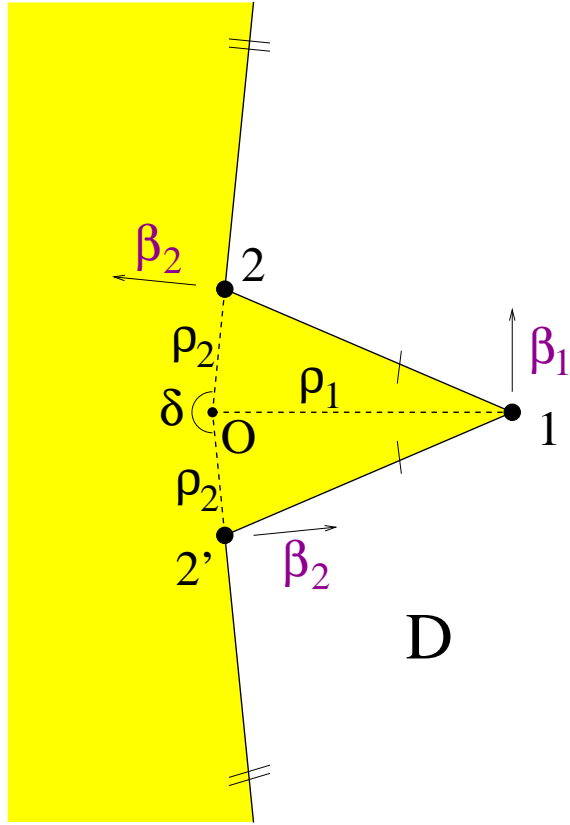}
\end{center}
\vskip 2 cm
\caption{%
A~spatial section of $D$ for a spinning spacetime at
$\sigma=0$. The figure shows the projection to the $(X,Y)$-plane and
suppresses the coordinate~$T$: if the plane of the paper is at 
$T=0$, then the corner labelled $1$ is in this plane, while
the corner labelled $2$ is above and the corner labelled $2'$ is below
this plane for $\beta>0$, and conversely for $\beta<0$. 
$\beta_i$, given by~(\ref{betaonetwo-def}), 
are the boost parameters of the particle trajectories with
respect to the coordinates $(T,X,Y)$: for 
$\beta>0$, $\beta_i$ are positive and
the velocities are in the directions shown, whereas for $\beta<0$, 
$\beta_i$ are negative and the directions are the
opposite. 
$\rho_1$~and $\rho_2$ are as given in~(\ref{rhos-def}). 
The single-stroked boundaries (neither of which is 
parallel to the plane of the paper) 
are identified by a
rotation in the spacetime about the world line of particle $1$ 
by the angle~$\delta_1$. 
The double-stroked boundaries
(each of which is parallel to the plane
of the paper but not in this plane) 
are identified by~(\ref{J-action}). 
Note that 
the origin of the $(X,Y)$ plane, labelled~$O$, is
outside~$D$. 
}
\label{fig:spinfundomain-closest}
\end{figure}

\begin{figure}
\begin{center}
\vglue 2 cm
\leavevmode
\epsfysize=12cm
\epsfbox{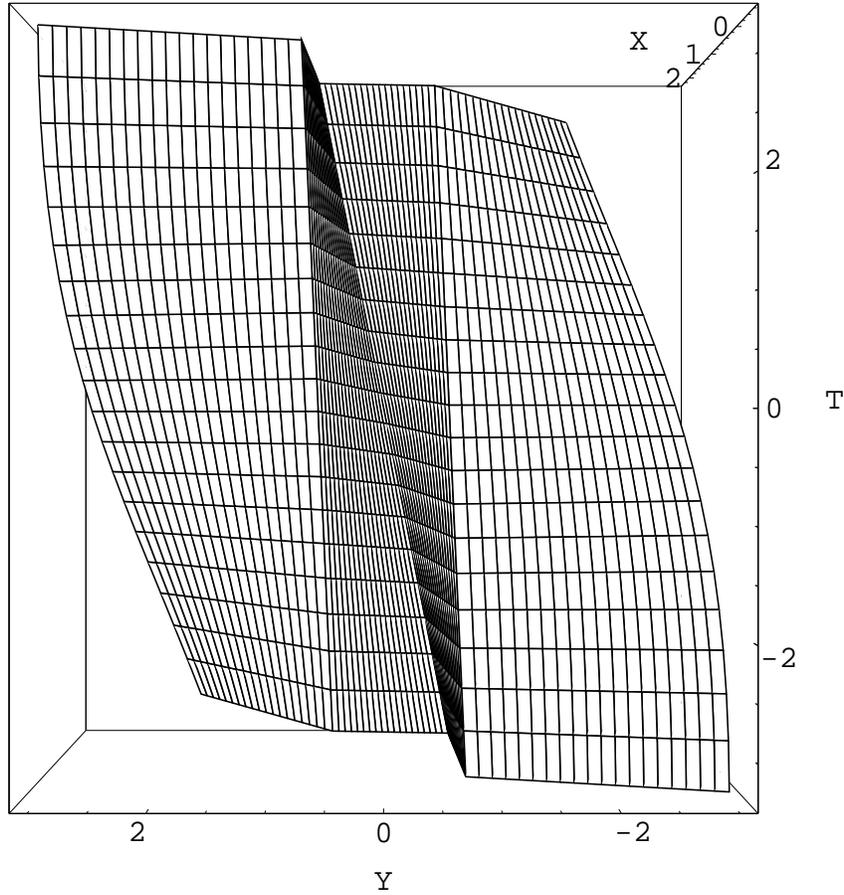}
\end{center}
\vskip 3 cm
\caption{%
The 
boundary of the fundamental domain $D$
for the spinning spacetime 
in the Minkowski coordinates $(T,X,Y)$. 
$D$~is behind the boundary, and the 
parameters are 
$\delta_1 = 2\pi/5$, $\delta_2 = 4\pi/5$, $a=1$, and $\beta = 0.2$. 
The viewpoint is on the negative $X$-axis. 
The grid is adapted to the
respective identifications of the boundaries 
as in figures \ref{fig:static3d-fundomain}
and~\ref{fig:coll3d-fundomain}. 
Note how the
identifications affect the Minkowski time coordinate~$T$.
}
\label{fig:spin3d-fundomain}
\end{figure}

Figure \ref{fig:spinfundomain-closest} shows the $\sigma=0$
configuration of $D$ in the 
the Minkowski coordinates 
$(T,X,Y)$, with the $T$-coordinate suppressed: 
the plane of the
paper is at $T=0$ and corner $1$ is in this plane, while
corner $2$ is above and corner $2'$ is below
this plane for $\beta>0$, and conversely for $\beta<0$. 
Figure 
\ref{fig:spinfundomain-later} shows the configuration at a larger
value of~$\sigma$, with the $T$-coordinate similarly suppressed: this
configuration is later (earlier) for $\beta>0$
($\beta<0$). 
A~spacetime picture of
$D$ is shown in
figure \ref{fig:spin3d-fundomain}
(for $\beta>0$). 
Note how all the boundary segments of 
$D$ twist as the particles evolve in~$T$: 
none of these segments is in a timelike plane. 

Near the infinity, we introduce conical coordinates
$(t,R,\varphi)$ as in subsection~\ref{subsec:colliding}, replacing in 
the transformation (\ref{transf}) 
$t$ by $t-t_0$ and 
$\varphi$ by $\varphi-\varphi_0$, and we extend the neighbourhood of
the infinity inwards to the particle world lines. 
In terms of the parameter~$\sigma$,
the particle trajectories in the conical coordinates are then 
given by (\ref{Rs-of-sigma}) with 
\begin{mathletters}
\label{varphis-of-sigma}
\begin{eqnarray}
&&\varphi_1 = \varphi_0 + \sigma/\alpha
\ \ ,
\label{varphione-of-sigma}
\\
&&\varphi_2 = \varphi_0 + \sigma/\alpha + \pi
\ \ ,
\end{eqnarray}
\end{mathletters}
\begin{equation}
t_1 = t_2 
= t_0 + 
a \left[
\frac{\cosh \beta_1 \cosh \beta_2 }{\sinh\beta} \tan\sigma
- \frac{s_1 s_2 \sinh\beta}{\pi\alpha\sin(\delta/2)} \sigma
\right]
\ \ .
\label{ts-of-sigma}
\end{equation}
The particles at a given $\sigma$ 
are thus on the {\em same\/} constant $t$ surface. 
Equation 
(\ref{ts-of-sigma}) can be uniquely inverted for $\sigma$ as a
function of $t\in(-\infty,\infty)$, 
and the resulting function $\sigma(t)$
vanishes at $t=t_0$. 
Substituting this $\sigma(t)$ in 
(\ref{Rs-of-sigma}) and (\ref{varphis-of-sigma}) 
yields the particle
trajectories 
in the form 
$\biglb(t,R_i(t),\varphi_i(t)\bigrb)$. 

Equations (\ref{varphis-of-sigma}) show 
that the particle world lines intersect a
constant $t$ surface at values of $\varphi$ that differ
by~$\pi$. In this sense, the particles are at each $t$ again 
diametrically opposite each other. 
To understand geometrically the 
constants $t_0$ and~$\varphi_0$, 
we observe that the length of the geodesic
connecting the particles at a given $t$ is 
\begin{equation}
R_c :=
\frac{a}{\cos\sigma}
\sqrt{1 +
\frac{s_1^2 s_2^2 \sinh^2\beta \sin^2\sigma}{\sin^2(\delta/2)}
} 
\ \ ,
\label{Rc-def}
\end{equation}
which reaches its minimum at $\sigma=0$, or in other words 
at $t=t_0$, at which
moment $\varphi_1 = \varphi_0$. 
$t_0$~is therefore the moment of conical time 
at which the particles 
are at their smallest spatial
separation, and $\varphi_0$ is
the conical angle of particle $1$ at this moment. 
The constants $t_0$ and $\varphi_0$ thus encode the 
zero-point of time and the orientation of the two-particle system
relative to the conical coordinates. 

It follows from the general considerations of 
\cite{gott1,ori,deser-jack-th-prl} 
that the spacetime does not have closed 
causal curves. In particular, it can be verified that $\rho_i$ 
are larger than the
critical radius $|S|/\alpha$ at which closed causal curves appear 
in the spacetime of a single
spinning particle. 

Finally, note that the static spacetimes of subsection
\ref{subsec:static} can be obtained from the spinning spacetimes
in the limit $\beta\to0$. 
The correct limiting forms 
for formulas involving the parameter 
$\sigma$ (which becomes ill-defined in the
limit) 
arise after first replacing 
$\sigma$ by the new parameter 
$\lambda:=a\sigma/\beta$, which always increases toward the future. 
Similarly, the colliding-particle spacetimes of subsection 
\ref{subsec:colliding} 
can be obtained
from the spinning spacetimes in 
the limit $a\to0$ with fixed~$\beta\ne0$, provided one first restricts
$t-t_0$ to having only one sign, positive (negative) 
values yielding a spacetime to 
the future (past) of the collision. 
If the sign of $t-t_0$ is unrestricted, the limit 
$a\to0$ with fixed $\beta\ne0$ is ambiguous: 
the reason is that for $S\ne0$, the particles scatter off each other
so that each 
conical angle changes from the asymptotic past to the asymptotic
future by
$(\pi/\alpha)\,{\rm sgn}(\beta)$, and the $a\to0$ limit of this
quantity with fixed $\beta\ne0$ depends on the sign of~$\beta$.
Constructing a colliding-particle spacetime that contains
both the past and the future of the collision requires thus additional 
assumptions: an example is the elastic collision discussed in 
\cite{head-gott}.

\section{Action in the connection formulation}
\label{sec:action-connection}

In this section we recall a first-order formulation of 2+1
gravity \cite{achu,witten1,AAbook2,romano} 
with massive point particles \cite{matsch-well}. 
We follow 
the notation of \cite{AAbook2,romano,loumar-wittentorus}, 
with the exception that we use units in which 
$4\pi G = 1$ \cite{matsch-well}.

\subsection{Bulk action}

The $(2+1)$-dimensional gravitational field in the connection
formulation is a connection in an $\IOctwoone$ bundle over the
three-dimensional spacetime manifold.  With our spacetime topology the
relevant bundle is the trivial one\footnote{Our spacetime is
  topologically 
  the product of a twice punctured plane and the real line, and the
  tangent bundle of this spacetime is trivial. A~nondegenerate triad
  provides a linear isomorphism between the tangent bundle and the
  bundle of local Lorentz frames.}, and we can without loss of
generality work in a global trivialisation.
The gravitational field can then be written as 
the globally-defined 
$\Octwoone$ connection
one-form 
${\bar A}_a^I$, taking values in the Lie algebra 
$\otwoone\simeq\sltwor$, and the globally-defined 
co-triad ${\bar 
e}_{aI}$ one-form, taking values in the dual of this Lie algebra. 
The internal
indices $I,J,\ldots$ take values in $\left\{0,1,2\right\}$, and they
are raised and lowered with the internal Minkowski metric,
$\eta_{IJ}={\rm{diag}}(-1,1,1)$. The indices $a,b,\ldots$ are abstract 
spacetime indices. 

The bulk action reads 
\begin{equation}
S 
= \case14 \int_{M} {d^3x} \,
\tilde{\eta}^{abc}
\, {\bar e}_{aI} {\bar F}^I_{bc}
\ \ ,
\label{3-action}
\end{equation}
where $\tilde{\eta}^{abc}$ is the Levi-Civita density and
${\bar F}^I_{bc}$ is the curvature of the connection,
\begin{equation}
{\bar F}^I_{bc} = 2 \partial^{\phantom{I}}_{[b} {\bar A}^I_{c]} +
\epsilon^I{}_{JK} {\bar A}^J_b {\bar A}^K_c
\ \ .
\end{equation}
The structure constants $\epsilon^I{}_{JK}$ are obtained from the totally
antisymmetric symbol $\epsilon_{IJK}$ by raising the index with the
Minkowski metric. Our convention is $\epsilon_{012}=1$. When the
co-triad is nondegenerate, the metric ${\bar e}_{aI} {\bar e}_{b}^I$
has signature 
$(-,+,+)$, and the field equations derived from
(\ref{3-action}) imply flatness of the metric,
which is equivalent to the metric's 
satisfying the vacuum Einstein equations. 

Changing the global trivialisation of the $\IOctwoone$ bundle gives 
rise to the gauge transformation 
\begin{mathletters}
\label{gaugetransae}
\begin{eqnarray}
{\bar A}^I_a K_I 
&\mapsto&
{\left(\Rmatrix^{-1}\right)}^I
{\vphantom{\left(\Rmatrix^{-1}\right)}}_J \,
{\bar A}^J_a K_I
+ \Rmatrix^{-1} \partial_a \Rmatrix
\ \ ,
\label{gaugetransa}
\\
{\bar e}^I_a 
&\mapsto&
{\left( \Rmatrix^{-1}\right)}^I
{\vphantom{\left(\Rmatrix^{-1}\right)}}_J
\left(
{\bar e}^J_a
+ {\bar{\cal D}}_a w^J
\right)
\ \ ,
\label{gaugetranse}
\end{eqnarray}
\end{mathletters}
where the matrix $\Rmatrix$ 
takes values in the defining representation 
of~$\Octwoone$, 
${\bar {\cal D}}_a$~is the gauge-covariant 
derivative determined by~${\bar A}^I_a$,
\begin{equation}
{\bar {\cal D}}_a v_K =
\partial_a v_K - \epsilon^I{}_{JK} {\bar A}^J_a v_I
\ \ , 
\label{dbar}
\end{equation}
and $K_I$ are the
$\otwoone$ adjoint representation 
basis matrices with the components ${\left(K_J\right)}^I
{\vphantom{\left(K_I\right)}}_L=\epsilon^I{}_{JL}$, 
\begin{equation}
K_0 = \pmatrix { 0 & 0 & 0 \cr 0 & 0 & -1 \cr 0 & 1 & 0 \cr } 
\ \ , 
\ \ 
K_1 = \pmatrix { 0 & 0 & -1 \cr 0 & 0 & 0 \cr -1 & 0 & 0 \cr } 
\ \ , 
\ \ 
K_2 = \pmatrix { 0 & 1 & 0 \cr 1 & 0 & 0 \cr 0 & 0 & 0 \cr } 
\ \ .
\end{equation}
We recall for future use the identities 
\begin{mathletters}
\label{K-identities}
\begin{eqnarray}
&&
\Rmatrix K_I \Rmatrix^{-1} = 
{\Rmatrix}^J
{\vphantom{\Rmatrix}}_I \, K_J
\ \ ,
\label{RK-identity}
\\
&&
{\rm Tr} (K_I K_J) = 2
\eta_{IJ}
\ \ .
\label{trace-ident}
\end{eqnarray}
\end{mathletters}
If the transformation (\ref{gaugetransae}) is connected to the
identity, it leaves the action (\ref{3-action}) invariant. If the 
transformation is not connected to the identity, the action
(\ref{3-action}) may acquire a topological additive constant.

We now take the spacetime manifold to be 
$\BbbR\times\Sigma$, where 
$\Sigma$ is the plane with two punctures. 
The 2+1 decomposition of the bulk action reads 
\cite{witten1,AAbook2,romano}
\begin{equation}
S_{\rm bulk} = \casehalf \int dt \int_\Sigma d^2x
\left[
{\tilde E}^j_I \left( \partial_t A^I_j \right) +
{\bar A}^I_t  {\cal D}_j {\tilde E}^j_I 
+ \casehalf {\bar e}_{tI} F^I_{ij} {\tilde \eta}^{ij}
\right]
\ \ .
\label{2+1-bulkaction}
\end{equation}
The abstract indices $i,j,\ldots$ live on~$\Sigma$, and $t$ is the
coordinate on~$\BbbR$. The $\Octwoone$ connection $A^I_j$ is
the pull-back of ${\bar A}_a^I$ to~$\Sigma$, $F^I_{ij}$ is its
curvature, 
given by
\begin{equation}
F^I_{ij} = 2 \partial^{\phantom{I}}_{[i} A^I_{j]} +
\epsilon^I{}_{JK} A^J_i A^K_j
\ \ ,
\end{equation}
and ${\tilde \eta}^{ij}$ is the Levi-Civita density on~$\Sigma$.
The vector density ${\tilde E}^j_I$ is given by ${\tilde E}^j_I={\tilde
\eta}^{ji} e_{iI}$, where $e_{iI}$ is the pull-back of ${\bar e}_{aI}$
to~$\Sigma$. ${\cal D}_j$~is the gauge-covariant derivative on $\Sigma$
determined by~$A^I_j$,
\begin{equation}
{\cal D}_j v_K =
\partial_j v_K - \epsilon^I{}_{JK} A^J_j v_I
\ \ .
\end{equation}
The canonical pair is thus $(A^I_j,{\tilde E}^{j}_{I})$,
and the Poisson brackets read 
\begin{equation}
\left\{A^I_i(x) , {\tilde E}^j_J (x')
\right\} = 2 \, \delta^j_i \delta^I_J \delta \left(x,x'\right)
\ \ ,
\label{AE-PB}
\end{equation}
where $x$ and $x'$ denote points on~$\Sigma$. 
${\bar e}_{tI}$~and ${\bar
A}^I_t$ act as Lagrange multipliers enforcing the constraints
\begin{mathletters}
\label{constraints}
\begin{eqnarray}
F^I_{ij} &=& 0
\label{F-constraint}
\\
{\cal D}_j {\tilde E}^j_I &=& 0
\ \ .
\label{G-constraint}
\end{eqnarray}
\end{mathletters}

\subsection{Boundary conditions and boundary terms}
\label{subsec:bc-bt}

We now turn to the boundary conditions. 
{}From now on we assume that the 
co-triad 
${\bar e}_{aI}$ is nondegenerate everywhere on~$\Sigma$. 
We write 
${\bbox{\bar A}}^I := {\bar A}_a^I dx^a$, 
${\bbox{\bar e}}^I := {\bar e}_a^I dx^a$, 
${\bbox{A}}^I := {A}_j^I dx^j$, 
${\bbox{e}}^I := {e}_j^I dx^j$. 

Near the infinity, we introduce on $\Sigma$ polar coordinates
$(r,\varphi)$, identified as $(r,\varphi)\sim (r,\varphi+2\pi)$,
such that the infinity is at $r\to\infty$. We assume that in some
neighbourhood of the infinity 
the variables take the form 
\begin{mathletters}
\label{conical-infty-falloff}
\begin{eqnarray}
{\bbox{\bar e}}^0 
&=& 
dt + S d\varphi
\ \ ,
\\
{\bbox{\bar e}}^1 
&=& 
dr
\ \ ,
\\ 
{\bbox{\bar e}}^2 
&=& 
\alpha r d\varphi
\ \ ,
\\
{\bbox{\bar A}}^0 
&=& 
\alpha d\varphi
\ \ ,
\\
{\bbox{\bar A}}^1 
&=& 
{\bbox{\bar A}}^2 =0
\ \ , 
\end{eqnarray}
\end{mathletters}
where $\alpha$ and
$S$ may depend on~$t$, and they satisfy 
$\alpha>0$, $-\infty<S<\infty$. The integral defining 
$S_{\rm bulk}$ is then convergent at the
infinity (since the integrand in (\ref{2+1-bulkaction}) vanishes when
(\ref{conical-infty-falloff}) holds), and the variation of 
$S_{\rm bulk}$
acquires
from the infinity the boundary term $-\pi \int dt \,
\delta\alpha$. This boundary term is cancelled 
provided we add to $S_{\rm bulk}$ the infinity boundary
action 
\begin{equation}
S_{\infty} := 
\pi \int dt \, (\alpha -1)
\ \ .
\label{infty-action}
\end{equation}
The constant term in the integrand in (\ref{infty-action}) has
been chosen for later convenience. 

The field equations for the ansatz (\ref{conical-infty-falloff}) are
equivalent to the $t$-independence of 
$\alpha$ and~$S$. When $\alpha$ and $S$ are $t$-independent, 
the metric obtained from (\ref{conical-infty-falloff}) is the 
conical metric of section~\ref{sec:spinning-point}, 
and $(t,r,\varphi)$ are a set of conical 
coordinates. The infinity behavior
(\ref{conical-infty-falloff}) and the boundary action
(\ref{infty-action}) therefore reproduce the desired classical
solutions near the infinity. 

Consider then the particles, which we label by 
the index $i=1,2$ as in
section~\ref{sec:spacetimes}.
We denote the masses by~$m_i$, we regard these masses as prescribed
parameters, and we assume $0<m_i<\pi$, $0<m_1 + m_2 <\pi$. 
Near each particle, we introduce on $\Sigma$ local polar coordinates 
$(r,\varphi)$, identified as $(r,\varphi)\sim (r,\varphi+2\pi)$, 
such that the particle is at the puncture of $\Sigma$ at 
$r\to0_+$. (We suppress on these
coordinates the index pertaining to the particle.) 
The boundary actions 
at the particles read 
\cite{matsch-well}, in our notation, 
\begin{equation}
S_{i} := 
\casehalf \int dt \int_{r=0} d\varphi \, 
{\bar A}^I_t e_{I\varphi} 
- \int dt \, \zeta_i \left[ \casehalf {\rm Tr}(\bbox{u}_i) - \cos (m_i)
\right]
\ \ , 
\label{action-particle-i}
\end{equation}
where 
$\zeta_i$~is a Lagrange
multiplier and $\bbox{u}_i$ is the $\SLtwor$-holonomy of 
$A^I_j$ around the particle, 
\begin{equation}
\bbox{u}_i := {\cal P} \exp 
\left[ \casehalf 
\int_{\varphi=0}^{\varphi=2\pi} 
d\varphi \, 
\left(A^I_\varphi\right)_{r=0} \gamma_I
\right]
\ \ ,
\end{equation}
where ${\cal P}\exp$ is the path-ordered exponential and the
$2\times2$ matrices $\gamma_I$
are a basis for $\sltwor$, 
\begin{equation}
\gamma_0 = \pmatrix { 0 & 1 \cr -1 & 0 } 
\ \ , 
\ \ 
\gamma_1 = \pmatrix { 0 & 1 \cr 1  & 0 }
\ \ , 
\ \ 
\gamma_2 = \pmatrix { 1 & 0 \cr 0 & -1 } 
\ \ . 
\end{equation}
$\bbox{u}_i$ depends on the choice of the coordinates 
$(r,\varphi)$ via $\SLtwor$ conjugation, corresponding to changing the
direction of $\varphi=0$, 
but as 
${\rm Tr}(\bbox{u}_i)$ is invariant under conjugation, 
$S_{i}$
(\ref{action-particle-i})
is independent of this choice. 
With the variation of ${\bar A}^I_t$ unrestricted, the
variation of the total action with respect to ${\bar A}^I_t$
then yields at $r=0$ the constraint $e^I_\varphi=0$: this means that 
the co-triad becomes degenerate in the limit $r\to0_+$ in such a way
that the proper circumference about $r=0$ vanishes. The 
variation with respect to $\zeta_i$ yields the constraint 
\begin{equation}
\casehalf {\rm Tr}(\bbox{u}_i) - \cos (m_i)
=0
\ \ .
\end{equation}
As discussed in \cite{matsch-well}, this implies that the extremal
geometry is near $r=0$ a spinless conical geometry whose defect
angle $\delta_i$ satisfies $\cos(\delta_i/2) = \cos (m_i)$. 
We require 
the defect angles to satisfy 
$0\le \delta_i < 2\pi$: this can be achieved 
by adopting near $r=0$ suitable falloff conditions whose detailed form 
is not important here. With these conditions, the boundary actions 
(\ref{action-particle-i}) therefore reproduce the desired classical
solutions near the particles. 

To summarise, the desired classical solutions are recovered by varying 
the action 
\begin{equation}
S_{\rm total} := S_{\rm bulk} + S_{\infty} + S_1 + S_2
\label{S-total}
\end{equation}
under our boundary conditions. 
The constraint algebra and the gauge transformations of 
$S_{\rm total}$ are discussed in 
\cite{matsch-well}.

\section{Hamiltonian reduction}
\label{sec:reduction}

In this section
we reduce the action by imposing
on the canonical pair $(A^I_j,{\tilde E}^{j}_{I})$ 
the constraints and fixing
the gauge. We take advantage of the explicit
knowledge of the classical solutions 
in the form given in section~\ref{sec:spacetimes}: 
restricting in this section 
the attention to the spinning spacetimes, 
we parametrise the initial data $(A^I_j,{\tilde E}^{j}_{I})$ 
in terms of the quadruple 
$(\beta,a,\varphi_0,\sigma)$, which specifies a spinning spacetime and 
a spacelike surface in it near the infinity, and we then show that
$(\beta,a,\varphi_0,\sigma)$ provides a (noncanonical) chart on the
reduced phase space and evaluate the symplectic structure. A~reader
not interested in the technicalities of the gauge-fixing conditions
and the evaluation of the reduced action may wish to proceed directly 
to equations (\ref{palpha-pvarphinougt-defs})--(\ref{capM-def}), 
which give the reduced action in terms of the quadruple 
$(\beta,a,\varphi_0,\sigma)$.

\subsection{Embedding of $\Sigma$ 
in a fictitious two-particle spacetime}
\label{subsec:embedding}

The constraints in $S_{\rm total}$ 
imply that the fields $(A^I_j,{\tilde
  E}^{j}_{I})$ on $\Sigma$ are induced by 
embedding $\Sigma$ in some (for the moment fictitious) 
two-particle Einstein spacetime of the form
discussed in section~\ref{sec:spacetimes}. 
We assume from now on that this embedding spacetime
has nonvanishing spin: the 
embedding spacetime is then 
specified up to isometries by the
pair $(\alpha,S)$ with $0<\alpha<1$ and $S\ne0$, or
equivalently by the pair 
$(\beta,a)$ with $\beta\ne0$ and $a>0$. 

\begin{figure}
\begin{center}
\vglue 3 cm
\leavevmode
\epsfysize=5cm
\epsfbox{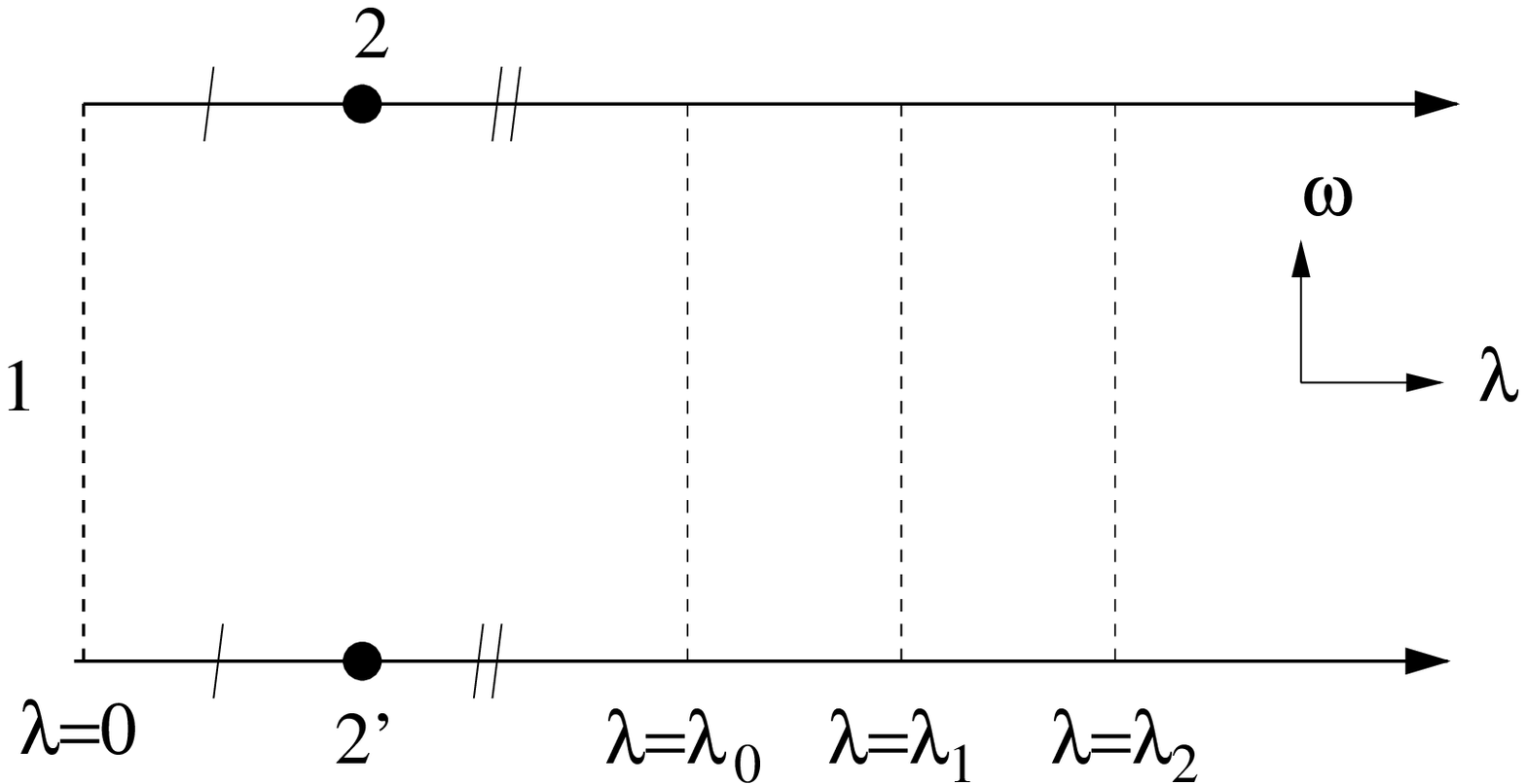}
\end{center}
\vskip 3 cm
\caption{%
The fundamental half-strip 
$\Omega := \left\{(\lambda,\omega) \mid \lambda>0, 
-\pi < \omega <\pi \right\}$
on the surface~$\Sigma$. 
On the boundary of~$\Omega$, particle $1$
is at the dashed line $\lambda=0$, 
while the points labelled $2$ and~$2'$,
respectively at $(\lambda,\omega) = (1,\pm\pi)$, are both at the second 
particle. The boundary segments indicated by a single (respectively 
double) stroke
are identified. 
The dashed lines $\lambda = \lambda_i$, $i\in\{0,1,2\}$, divide
$\Omega$ into regions in which the gauge choice is as explained in the 
text. 
}
\label{fig:corot-planardomain}
\end{figure}

We introduce on $\Sigma$ the simply-connected fundamental
region $\Omega$ coordinatised by the pair $(\lambda,\omega)$ as shown
in figure~\ref{fig:corot-planardomain}: 
$\Omega := \left\{(\lambda,\omega) \mid
  \lambda>0, -\pi < \omega <\pi \right\}$. 
The boundaries of $\Omega$ at $\omega = \pm\pi$
are identified as 
$(\lambda,\omega)\sim (\lambda,\omega+2\pi)$.
Particle $1$ is 
on the boundary of $\Omega$ at the line $\lambda=0$,
while the second particle is on the boundary of $\Omega$ 
at the two points labelled 
$2$ and~$2'$, respectively at 
$(\lambda,\omega) =
(1,\pm\pi)$. 

In the neighbourhood of the infinity, 
the embedding of $\Sigma$ is by construction in a spacelike surface of 
constant conical time. We specify this surface by the quadruple 
$(\beta,a,\varphi_0,\sigma)$. 
We wish to specify the embedding so that near the infinity 
$(\lambda,\omega)$ are the spatial conical coordinates of the 
embedding spacetime, while near the particles $(\lambda,\omega)$ 
are suitably adapted to the motion of the particles. 

To achieve this, we introduce the three numbers $\lambda_0$,
$\lambda_1$, and $\lambda_2$, satisfying 
$\lambda_0<\lambda_1<\lambda_2$, such that $\lambda_0$ is greater than 
the conical radii
(\ref{Rs-of-sigma}) of the particles in the
embedding spacetime at this conical time. For technical convenience,
we may assume $\lambda_0>1$. We now specify the embedding separately
in the regions 
$0 < \lambda \le \lambda_0$, 
$\lambda_0 \le \lambda \le \lambda_1$, 
$\lambda_1 \le \lambda \le \lambda_2$, 
and 
$\lambda \ge \lambda_2$ 
(see figure~\ref{fig:corot-planardomain}). 

Throughout the region 
$\lambda \ge \lambda_0$, 
we take the embedding to be
in the surface of constant conical time, 
and in this region we relate $(\lambda, \omega)$
to the spatial conical coordinates $(R,\varphi)$ by 
\begin{mathletters}
\label{corot-transf}
\begin{eqnarray}
R 
&=& 
\lambda
\ \ ,
\\
\varphi 
&=&
\omega + (\varphi_0 +
\sigma/\alpha)h(\lambda)
\ \ ,
\end{eqnarray}
\end{mathletters}
where 
$h(\lambda)$ is a smooth monotonic function
satisfying 
\begin{equation}
h(\lambda) = 
\cases{1\ , & $\lambda\le\lambda_1$ \ \ , \cr
0\ , & $\lambda\ge\lambda_2$ \ \ .\cr}
\end{equation}
For $\lambda\ge\lambda_2$, the coordinates $(\lambda, \omega)$ are
then the spatial 
conical coordinates of the embedding spacetime, as desired. For 
$\lambda_0 \le \lambda \le \lambda_1$, $\lambda$ is still the conical
radius, but $\omega$ has been made to co-rotate with the particles so
that $\omega$ vanishes at the conical angle of
particle~$1$, 
$\varphi = \varphi_0 +
\sigma/\alpha$. 
The interpolation between the conical coordinates and
the co-rotating coordinates takes place in the intermediate 
region,  
$\lambda_1 \le \lambda \le \lambda_2$. 

\begin{figure}
\begin{center}
\vglue 2 cm
\leavevmode
\epsfysize=10cm
\epsfbox{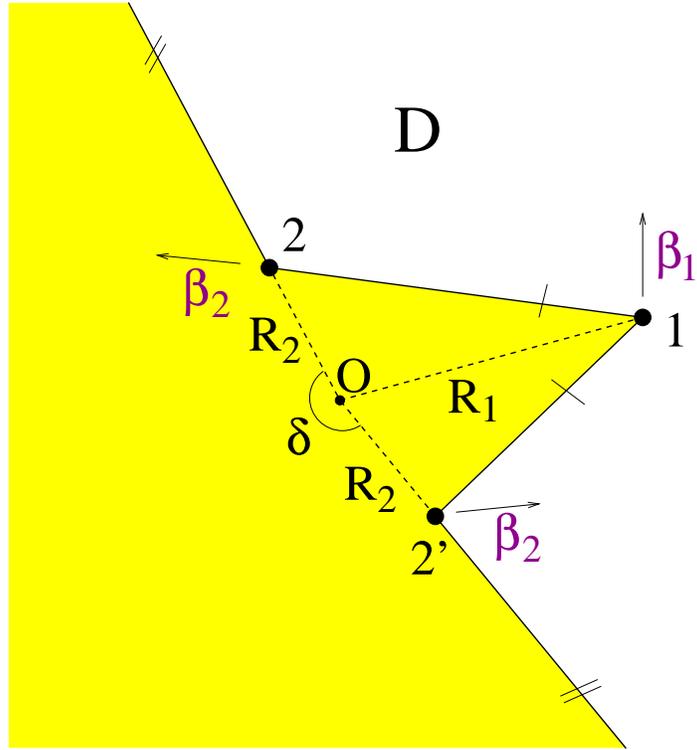}
\end{center}
\vskip 2 cm
\caption{%
A~spatial section of $D$ for a spinning spacetime at
$\sigma>0$. 
The coordinate $T$ is suppressed as in
figure~\ref{fig:spinfundomain-closest}. 
The single-stroked (double-stroked) boundaries 
are identified as in figure~\ref{fig:spinfundomain-closest}. 
$R_1$~and $R_2$ are given by~(\ref{Rs-of-sigma}). 
}
\label{fig:spinfundomain-later}
\end{figure}

The remaining and most technical part is 
to specify the embedding for 
$0 < \lambda \le \lambda_0$. Recall from 
subsection \ref{subsec:spinning} the
embedding of the fictitious two-particle spacetime into the Minkowski
fundamental region~$D$. 
In terms of this embedding, the boundaries of $\Omega$ at
$\lambda_0 \le \lambda \le \lambda_1$ 
lie at the boundaries of~$D$, and they
further lie on the 
spacelike section of $D$ 
shown in figure \ref{fig:spinfundomain-later}, on the
boundaries indicated there by double strokes. 
We now use this embedding
to specify the 
boundaries of $\Omega$ everywhere at $\lambda < \lambda_1$ (and thus
in particular at $\lambda < \lambda_0$): 
the stroked
and double-stroked boundaries of $\Omega$ in figure 
\ref{fig:corot-planardomain} are taken to be
at the corresponding 
stroked and double-stroked boundaries of the 
spacelike section of 
$D$ shown in figure~\ref{fig:spinfundomain-later}. On the 
double-stroked boundaries of~$\Omega$, 
we set 
\begin{mathletters}
\label{vectors-doublestroke}
\begin{eqnarray}
\partial_\lambda 
&=& 
f(\lambda)
\left[ 
\cos(\sigma \pm \alpha\pi) \partial_X 
+ \sin(\sigma \pm \alpha\pi) \partial_Y 
\right]
\ \ ,
\\
\partial_\omega 
&=& S \partial_T + 
\alpha \left[ 
- \sin(\sigma \pm \alpha\pi) \partial_X 
+ \cos(\sigma \pm \alpha\pi) \partial_Y 
\right]
\ \ , 
\end{eqnarray}
\end{mathletters}
where the upper (lower) sign pertains to the 
boundary component from $2$ ($2'$) towards increasing~$\lambda$. 
Here $f(\lambda)$ is a positive function that is equal to $1$ 
for $\lambda_0 \le \lambda \le \lambda_1$ and whose detailed form for 
$1\le \lambda \le \lambda_0$ 
is not important: it is introduced to account for the fact that
the points 
$2$ and $2'$ are at $\lambda=1$ but $R=R_2$ 
(\ref{Rtwos-of-sigma}). 
On the single-stroked boundaries~$\Omega$, 
we set 
\begin{mathletters}
\label{vectors-singlestrokes}
\begin{eqnarray}
\partial_\lambda 
&=& 
\pm \pi S \partial_T 
+ 
\left[ R_2 \cos(\sigma \pm \alpha\pi ) - R_1 \cos\sigma \right]
\partial_X 
+ 
\left[ R_2 \sin(\sigma \pm \alpha\pi ) - R_1 \sin\sigma \right]
\partial_Y
\ \ ,
\label{parlambda-singlestrokes}
\\
\partial_\omega &=& 
- \sin\psi_\pm \, \partial_X
+ 
\cos\psi_\pm
\left[ \sinh\beta_1 \, \partial_T + \cosh\beta_1 \, \partial_Y
\right]
\ \ , 
\label{paromega-singlestrokes}
\end{eqnarray}
\end{mathletters}
where the upper (lower) signs pertain to the boundary component
between $1$ and $2$~($2'$), and the angles $\psi_\pm$ are defined by 
\begin{mathletters}
\label{psiplusminus}
\begin{eqnarray}
\cos\psi_\pm 
&=& 
\frac{-c_1 \mp s_1 \cosh\beta_1 \, \tan\sigma}
{\sqrt{1 + \cosh^2 \beta_1 \, \tan^2\sigma}}
\ \ ,
\\
\sin\psi_\pm
&=& 
\frac{\pm s_1 - c_1 \cosh\beta_1 \, \tan\sigma}
{\sqrt{1 + \cosh^2 \beta_1 \, \tan^2\sigma}}
\ \ . 
\end{eqnarray}
\end{mathletters}
Equations (\ref{rhos-def})--(\ref{S-def}) show that 
$\partial_\lambda$ (\ref{parlambda-singlestrokes}) 
is the tangent vector to the affinely
parametrised spacelike
geodesic from $1$ to $2$~($2'$). 
$\partial_\omega$~(\ref{paromega-singlestrokes}) has been 
determined from 
the conditions that it is orthogonal to $\partial_\lambda$ and to the
vector $v_1 := \cosh\beta_1 \, \partial_T + \sinh\beta_1 \,
\partial_Y$ (which is the velocity of particle $1$ in~$D$), 
and pointing outward
(inward) on the boundary from $1$ to $2$~($2'$). 
We note for future use the decomposition 
\begin{equation}
\partial_\lambda = 
\left( a c_1 \sinh\beta_1 \, \tan\sigma \right)
v_1 
+ 
a {\left(1 + \cosh^2 \beta_1 \, \tan^2\sigma\right)}^{1/2}
\, 
u_\pm
\ \ ,
\label{parlambda-decomp}
\end{equation} 
where the spacelike unit vector $u_\pm$ is given by 
\begin{equation}
u_\pm := \cos\psi_\pm \, \partial_X
+ 
\sin\psi_\pm
\left[ \sinh\beta_1 \, \partial_T + \cosh\beta_1 \, \partial_Y
\right]
\ \ . 
\end{equation}
The three vectors $(v_1, u_\pm, \partial_\omega)$ form thus a 
Lorentz-orthonormal 
triad adapted to the velocity of particle $1$ and to the relative
positions of the points $1$ and $2$ (respectively~$2'$): $u_\pm$~is
obtained by rotating $\partial_X$ about $v_1$ by the angle~$\psi_\pm$, 
and $\partial_\omega$ is obtained by rotating 
$\partial_X$ about $v_1$ by the angle~$\psi_\pm+\pi/2$. 
Note that 
(\ref{psiplusminus}) implies 
$\psi_+ = \psi_- - \delta_1$  mod~$2\pi$, 
which must be the case by the construction of~$D$. 

It now follows from the identifications of the boundaries of 
$D$ that our embedding of $\Sigma$ in the fictitious
two-particle spacetime is $C^1$ across the identified boundaries
of~$\Omega$, and in particular the vectors 
$\partial_\lambda$ and $\partial_\omega$ are
continuous everywhere on~$\Sigma$. The embedding is smooth for
$\lambda\ge\lambda_0$, and it can clearly 
be chosen 
smooth everywhere by introducing suitable additional conditions, and we now
consider this done. Note that 
the embedding cannot be extended smoothly to the
boundary of $\Omega$ at $\lambda=0$ and at the points $2$ and~$2'$,
where the particles are. 
Note also that we have not specified the details of the embedding in
the interior of~$\Omega$: as will be seen in 
subsection~\ref{subsec:red-action}, 
these details will not be needed.

\subsection{Gauge choice}
\label{subsec:gauge-choice}

We now use the embedding of $\Sigma$ in the fictitious two-particle
spacetime to choose a gauge for the fields 
$(A^I_j,{\tilde E}^{j}_{I})$. 

Consider first the region $\lambda > \lambda_0$ of $\Omega$. We 
denote this 
region 
by~$\Omega_0$. 
Near
the infinity, the fields take
the form~(\ref{conical-infty-falloff}): when the
parameters in (\ref{conical-infty-falloff})
are time-independent, these fields
solve the field equations for 
$r$ larger than the conical radii of the
particles, and the coordinates 
in (\ref{conical-infty-falloff})
are directly the conical coordinates. 
We therefore adopt in $\Omega_0$ a gauge
by transforming the spatial projection of 
(\ref{conical-infty-falloff}) to the 
coordinates $(\lambda,\omega)$ 
by~(\ref{corot-transf}). The result is 
\begin{mathletters}
\label{outer-gauge}
\begin{eqnarray}
{\bbox{e}}^0 
&=& 
S \left[ 
d\omega + (\varphi_0 +
\sigma/\alpha)h' d\lambda
\right]
\ \ ,
\\
{\bbox{e}}^1 
&=& 
d\lambda
\ \ ,
\\ 
{\bbox{e}}^2 
&=& 
\alpha \lambda 
\left[ 
d\omega + (\varphi_0 +
\sigma/\alpha)h' d\lambda
\right]
\ \ ,
\\
{\bbox{A}}^0 
&=& 
\alpha 
\left[ 
d\omega + (\varphi_0 +
\sigma/\alpha)h' d\lambda
\right]
\ \ ,
\\
{\bbox{A}}^1 
&=& 
{\bbox{A}}^2 =0
\ \ , 
\end{eqnarray}
\end{mathletters}
where $h' := dh(\lambda)/d\lambda$. 

Consider then the region $\lambda < \lambda_1$ of $\Omega$. We 
denote this 
region 
by~$\Omega_1$. (Note that $\Omega_0$ and $\Omega_1$ overlap at
$\lambda_0 < \lambda < \lambda_1$.) 
We introduce on the 
fundamental domain $D$ of the fictitious spacetime 
the fields 
\begin{mathletters}
\label{flat-eA}
\begin{eqnarray}
{}_0{\bbox{\bar e}}^0 
&=& 
dT 
\ \ ,
\\
{}_0{\bbox{\bar e}}^1 
&=& 
dX
\ \ ,
\\ 
{}_0{\bbox{\bar e}}^2 
&=& 
d Y
\ \ ,
\\
{}_0{\bbox{\bar A}}^I
&=& 
0 
\ \ , 
\end{eqnarray}
\end{mathletters}
which satisfy the field equations and produce on $D$ the
Minkowski 
metric $ds^2 = - dT^2 + dX^2 + dY^2$. 
Let 
$({}_0A^I_j,{}_0{\tilde
  E}^{j}_{I})$ denote 
the fields obtained by the pull-back of (\ref{flat-eA})
to~$\Omega_1$. 
In order to obtain on $\Omega_1$ fields 
that can be continued to~$\Sigma$ and agree with 
(\ref{outer-gauge}) in the intersection of 
$\Omega_0$ and~$\Omega_1$, 
we perform on 
$({}_0A^I_j,{}_0{\tilde E}^{j}_{I})$ 
a (local) gauge transformation of the form 
(\ref{gaugetransae}) with $w^I=0$ and a judiciously-chosen~$\Rmatrix$: 

(i) 
For~$\lambda_0\le\lambda\le\lambda_1$, we take 
\begin{equation}
\Rmatrix = \exp[(\sigma + \alpha\omega) K_0]
\ \ . 
\label{Rgauge-infty}
\end{equation}
The resulting fields clearly agree with~(\ref{outer-gauge}). 
We further take (\ref{Rgauge-infty}) to hold everywhere near and on
the double-stroked
boundary components of $\Omega_1$ in
figure~\ref{fig:corot-planardomain}. 

(ii) 
Near and at $\lambda=0$, we take
\begin{equation}
\Rmatrix 
= 
\exp(- \beta_1 K_1) \exp(\psi_- K_0) 
\exp\{(\omega+\pi)[1-\delta_1/(2\pi)] K_0 \}
\ \ , 
\label{Rgauge-lambdazero}
\end{equation}
and we further 
take (\ref{Rgauge-lambdazero}) to hold everywhere near and on
the single-stroked
boundary components of~$\Omega_1$. This implies that on the
single-stroked boundary components themselves we have 
\begin{eqnarray}
\Rmatrix 
&=& 
\exp(- \beta_1 K_1) \exp(\psi_\pm K_0) 
\nonumber
\\
&=& 
\exp(\psi_\pm v_1^I  K_I) \exp(- \beta_1 K_1)
\ \ ,
\label{Rgauge-onestroke}
\end{eqnarray}
where the signs correspond to those
in~(\ref{vectors-singlestrokes}). The first equality in 
(\ref{Rgauge-onestroke}) follows using 
$\psi_+ = \psi_- - \delta_1$  mod~$2\pi$, 
and the second one using
$v_1^I = 
{\biglb( \exp(- \beta_1 K_1) \bigrb)}^I
{\vphantom{{\biglb( \exp(- \beta_1 K_1) \bigrb)}}}_0$
and~(\ref{RK-identity}). 

(iii) 
At the points $2$ and $2'$ on the boundary of~$\Omega_1$, 
$\Rmatrix$~cannot be defined consistently with 
both (\ref{Rgauge-infty}) and~(\ref{Rgauge-onestroke}). It will
suffice to assume that $\Rmatrix$ smoothly interpolates
between these boundary values
on (the
interior of)~$\Omega$. 

We claim that the resulting fields can be extended from $\Omega$
to~$\Sigma$.  For $\Omega_0$ this is obvious, and we only need to
consider~$\Omega_1$.

Consider $A^I_j$ in~$\Omega_1$. 
Where (\ref{Rgauge-infty}) holds, 
the only nonvanishing component of $A^I_j$ is $A^0_\omega = \alpha$, 
which is smooth across the
identification of the double-stroked boundaries. 
Where (\ref{Rgauge-lambdazero}) holds, 
the only nonvanishing component of $A^I_j$ is $A^0_\omega =
1-\delta_1/(2\pi)$, 
which is smooth across the
identification of the single-stroked boundaries. Thus $A^I_j$ extends
smoothly from $\Omega_1$ to~$\Sigma$. 

Consider then ${\tilde E}^{j}_{I}$ in~$\Omega_1$. 
On the double-stroked
boundaries, where 
(\ref{vectors-doublestroke}) and 
(\ref{Rgauge-infty}) hold, we have 
\begin{mathletters}
\label{infty-e-lambdaomega}
\begin{eqnarray}
{\bbox{e}}^0 
&=& 
S d\omega
\ \ ,
\\
{\bbox{e}}^1 
&=& 
f(\lambda) d\lambda
\ \ ,
\\ 
{\bbox{e}}^2 
&=& 
\alpha d\omega
\ \ ,
\end{eqnarray}
\end{mathletters}
which shows that $e^I_j$ is 
continuous on the identification of the 
double-stroked boundaries. An analogous calculation shows that
$e^I_j$ is continuous on the identification of the single-stroked 
boundaries: as seen from the last expression
in~(\ref{Rgauge-onestroke}), $\Rmatrix$~is precisely the matrix that
relates the orthonormal Minkowski triad $(\partial_T, \partial_X,
\partial_Y)$ to the orthonormal triad 
$(v_1, u_\pm, \partial_\omega)$ adapted to the double-stroked
boundaries of~$\Omega_1$, and 
the gauge transformation acting on the
internal index of 
${}_0{\bar e}^I_a$ (\ref{flat-eA}) matches 
on these boundaries 
the projection of 
the spacetime index $a$ to the spatial index
$j\in\{\lambda,\omega\}$. Thus $e^I_j$ and ${\tilde
E}^{j}_{I}$ extend continuously from $\Omega_1$ to~$\Sigma$. The 
extension can be chosen smooth by making further assumptions about the 
embedding of $\Sigma$ in~$D$, 
and we now consider this done. 

To summarise, we have obtained on $\Sigma$ fields 
$(A^I_j,{\tilde E}^{j}_{I})$ that satisfy the constraints. 
The gauge
has not been specified everywhere on~$\Sigma$, 
but it has been specified on and near the boundaries 
of the fundamental 
domain~$\Omega$, and 
the only parameters in this specification are 
$(\beta,a,\varphi_0,\sigma)$. 
We shall see that this is sufficient for 
evaluating the reduced action.

\subsection{Reduced action}
\label{subsec:red-action}

As all the constraints have been solved, the only terms remaining 
in $S_{\rm total}$ (\ref{S-total}) are $S_{\infty}$ and the
Liouville term of~$S_{\rm bulk}$. We now evaluate these terms. 

The parameters $(\beta,a,\varphi_0,\sigma)$ 
in our gauge fixing refer to a fictitious two-particle Einstein
spacetime, and to a spacelike surface in this
spacetime. We now interpret these
parameters as coordinates on the reduced phase space. When evaluating
the reduced action, all the parameters $(\beta,a,\varphi_0,\sigma)$
are then regarded as functions of~$t$.

Evaluating $S_{\infty}$~is immediate: 
the expression is as given in~(\ref{infty-action}), 
with $\alpha$ understood a function of
$\beta$ through (\ref{delta-deter}) and~(\ref{alphadelta-note}). 

In the Liouville term in~$S_{\rm bulk}$, 
the integral over
the region $\lambda>\lambda_1$ of $\Omega$ is straightforward 
using~(\ref{outer-gauge}), and 
yields to the Lagrangian the contribution 
$\pi S\left( 
\dot\sigma + \alpha {\dot \varphi}_0 - 
{\dot\alpha} \alpha^{-1}\sigma 
\right)$. 
In the region 
$\lambda<\lambda_1$~($\Omega_1$), a short calculation using 
(\ref{K-identities}) yields 
\begin{equation}
\casehalf 
{\tilde E}^j_I \left( \partial_t A^I_j \right)
= 
\case{1}{4}
\partial_j \left[
{\tilde E}^{jI} \, 
{\rm Tr}
\left(K_I \Rmatrix^{-1} {\dot \Rmatrix} \right) \right]
\ \ ,
\label{liouville-totalder}
\end{equation}
and the integral of (\ref{liouville-totalder})
over $\Omega_1$ can thus be converted into an 
integral over the boundary of~$\Omega_1$. 
We now consider the parts of 
this boundary in turn. 

On the boundary of $\Omega_1$ at $\lambda=\lambda_1$, 
$\Rmatrix$~is given by~(\ref{Rgauge-infty}), and from
(\ref{infty-e-lambdaomega}) we have 
${\tilde E}^{\lambda 0} = e^0_\omega = S$. 
Hence the contribution to
the Lagrangian is 
$- \pi S {\dot \sigma}$. 

The double-stroked boundary components of $\Omega_1$ 
are at $\omega = \pm\pi$,
$0 < \lambda< \lambda_1$. 
$\Rmatrix$~is given by~(\ref{Rgauge-infty}), and 
$\Rmatrix^{-1} {\dot \Rmatrix}$ is proportional to~$K_0$, but
(\ref{infty-e-lambdaomega}) implies ${\tilde E}^{\omega 0}=0$. The
contributions to the Lagrangian therefore vanish. 

On the boundary component at $\lambda=0$, the relevant components of 
${\tilde E}^{j I}$ are 
${\tilde E}^{\lambda I} = 
e^I_\omega$, and these vanish by our discussion of the particle
action $S_i$ (\ref{action-particle-i}) in
subsection~\ref{subsec:bc-bt}. As $\Rmatrix$ is regular, the
contribution to the Lagrangian
vanishes. A~similar argument applied to 
small half-circles about the singular points $2$ and $2'$
shows that the Lagrangian gets no contribution
from these singular points. 

What remains are the single-stroked boundary components of~$\Omega_1$,
at $\omega = \pm\pi$, $0<\lambda<1$. Their contribution
to the Lagrangian is 
\begin{equation}
\frac{1}{4}
{\left[ 
{\tilde E}^{\omega I} 
{\rm Tr}
\left(K_I \Rmatrix^{-1} {\dot \Rmatrix} \right)
\right]}_+
- 
\frac{1}{4}
{\left[ 
{\tilde E}^{\omega I} 
{\rm Tr}
\left(K_I \Rmatrix^{-1} {\dot \Rmatrix} \right)
\right]}_-
\ \ ,
\label{singles-raw}
\end{equation}
where the subscript $\pm$ indicates the component at 
$\omega = \pm\pi$. 
${\tilde E}^{\omega I}$~in (\ref{singles-raw}) can be written as 
\begin{eqnarray}
{\left( {\tilde E}^{\omega I} \right)}_\pm 
&=& 
- 
{\left[ 
{\left(\Rmatrix^{-1}\right)}^I
{\vphantom{\left(\Rmatrix^{-1}\right)}}_J 
\, {}_0 e^J_\lambda
\right]}_\pm 
\nonumber
\\
&=& 
- 
a c_1 \sinh\beta_1 \, \tan\sigma 
\pmatrix{ 
1 \cr
0 \cr
0 \cr}^{\!\!\!I}
- 
a  
{\sqrt{1 + \cosh^2 \beta_1 \, \tan^2\sigma}}
\pmatrix{ 
0 \cr
1 \cr
0 \cr}^{\!\!\!I}
\ \ ,
\label{Etilde-on-singles}
\end{eqnarray}
where the first equality follows from~(\ref{gaugetranse}), and the
second one from~(\ref{parlambda-decomp}), 
(\ref{flat-eA}), and~(\ref{Rgauge-onestroke}). As the last 
expression in (\ref{Etilde-on-singles}) is independent of the
subscript~$\pm$, ${\tilde E}^{\omega I}$ 
factors out in~(\ref{singles-raw}). In the remaining factor
in~(\ref{singles-raw}), we use 
the first expression in (\ref{Rgauge-onestroke}) to obtain 
\begin{eqnarray}
\Rmatrix^{-1}_+ {\dot \Rmatrix}_+ 
- 
\Rmatrix^{-1}_- {\dot \Rmatrix}_- 
&=& 
- {\dot\beta}_1 
\left[ 
\exp(-\psi_+ K_0) K_1 \exp(\psi_+ K_0)
- 
\exp(-\psi_- K_0) K_1 \exp(\psi_- K_0)
\right]
\nonumber
\\
&=& 
- {\dot\beta}_1 
{\left[ \exp(-\psi_+ K_0) - \exp(-\psi_- K_0) \right]}^J
{\vphantom{ {\left[ \exp(-\psi_+ K_0) - \exp(-\psi_- K_0) \right]}}}_1
K_J
\ \ .
\end{eqnarray}
where the last equality follows from~(\ref{RK-identity}). 
Using (\ref{trace-ident}) and~(\ref{psiplusminus}), 
we thus find that
(\ref{singles-raw}) is equal to 
\begin{equation}
- a s_1 {\dot\beta}_1  \cosh\beta_1 \tan\sigma
= 
\frac{ {\dot\alpha} \pi a 
\cosh\beta_1 \cosh\beta_2}{\sinh\beta}
\tan\sigma
\ \ ,
\label{singles-ripe}
\end{equation}
where the last equality follows using 
(\ref{betaone-def}) and the identity 
\begin{equation}
\frac{d}{d\alpha} \!\! 
\left[ 
\frac{\sinh\beta}{\sin(\delta/2)}
\right] 
= 
- \frac{\pi \cosh\beta_1 \cosh\beta_2}{s_1 s_2 \sinh\beta}
\ \ .
\end{equation}

Collecting, and defining 
\begin{mathletters}
\label{palpha-pvarphinougt-defs}
\begin{eqnarray}
&& 
p_\alpha := 
\pi a 
\left[
\frac{\cosh \beta_1 \cosh \beta_2 }{\sinh\beta} \tan\sigma
- \frac{s_1 s_2 \sinh\beta}{\pi\alpha\sin(\delta/2)} \sigma
\right]
\ \ ,
\\
&& 
p_{\varphi_0}:= 
\pi\alpha S
\ \ , 
\end{eqnarray}
\end{mathletters}
we find the reduced action 
\begin{equation}
S_{\rm red} = \int dt 
\left(
p_\alpha {\dot \alpha} 
+ p_{\varphi_0} {\dot\varphi}_0
- M
\right)
\ \ ,
\label{S-alphavarphinought}
\end{equation}
where 
\begin{equation}
M:=\delta/2 = \pi(1-\alpha)
\ \ . 
\label{capM-def}
\end{equation}
The quadruple $(\beta,a,\varphi_0,\sigma)$, with $\beta\ne0$ and
$a>0$, therefore provides a (noncanonical) 
chart on the reduced phase space, as
promised. This chart consists of two disjoint patches, one with 
$\beta>0$ and the other with $\beta<0$. We 
denote the reduced phase
space covered by this chart by~$\Gammaredplus$. 
A~canonical chart on $\Gammaredplus$ is provided by 
$(\alpha,\varphi_0 ; p_\alpha,p_{\varphi_0})$: 
this chart consists of the two disjoint patches
$p_{\varphi_0}>0$ and $p_{\varphi_0}<0$, 
and in each patch 
$p_\alpha$ takes all real values, 
$\varphi_0$ takes all real values 
modulo~$2\pi$, and $0<\alpha < 1 - (m_1 + m_2)/\pi$. 
Note that the Hamiltonian~$M$, given by (\ref{capM-def}), arose
from the infinity boundary action $S_{\infty}$~(\ref{infty-action}). 

It is easily verified that the action
(\ref{S-alphavarphinought}) on $\Gammaredplus$ 
correctly reproduces the classical
solutions with $S\ne0$. 
What is missing, however, are the classical solutions with 
$S=0$. 
We shall
obtain an action from which also the solutions with $S=0$ can be
recovered in section~\ref{sec:com-coords}.

\section{New phase space chart: 
`Configuration' and `momentum' 
at a conical time}
\label{sec:com-coords}

The canonical chart $(\alpha,\varphi_0 ; p_\alpha,p_{\varphi_0})$
on~$\Gammaredplus$ is adapted to the spacetime properties of the
spinning classical solutions. We now introduce on
$\Gammaredplus$ a canonical chart in which the variables 
reflect more closely the geometrical 
`configuration' of the two particles at a
moment of conical time.

Recall that the spatial geodesic distance of the particles at a 
moment of conical time is $R_c$~(\ref{Rc-def}). Recall also that the
conical angles of the particles differ by~$\pi$, so that the orientation
of the particles 
with respect to the infinity is completely 
specified by (say) the conical angle of 
particle $1$ (\ref{varphione-of-sigma}). 
We relabel this angle as~$\varphi_c$: 
\begin{equation}
\varphi_c :=
\varphi_0 + \sigma/\alpha
\ \ .
\label{varphic-def}
\end{equation}
Geometrically, the pair $(R_c,\varphi_c)$ then characterises a
`configuration' of the particles with respect to the infinity
at a moment of conical time.  Further, $R_c$~and $\varphi_c$ Poisson
commute. 

Define now on $\Gammaredplus$ the functions 
\begin{mathletters}
\label{transf-redmoms}
\begin{eqnarray}
&&P_c :=
\arcsinh 
\left(
\frac{s_1 s_2 \sinh\beta \sin\sigma}{\sin(\delta/2)}
\right)
\ \ ,
\label{transf-redmom-Pc}
\\
&&p_{\varphi_c} :=
p_{\varphi_0}
\ \ .
\end{eqnarray}
\end{mathletters}
It is tedious but elementary to verify that the quadruple 
$(R_c,\varphi_c;P_c,p_{\varphi_c})$ provides a new two-patched chart
on~$\Gammaredplus$, such that the ranges of the coordinate functions are 
$p_{\varphi_c}\ne0$, $R_c>0$, and 
\begin{equation}
\frac{\pi |p_{\varphi_c}| \cosh(P_c)}{R_c} < 
c_1 + c_2
\ \ .
\label{cm-chart-ineq}
\end{equation}
The action in the new chart reads 
\begin{equation}
S_{\rm red} = \int dt 
\left( 
P_c {\dot {R_c}} + p_{\varphi_c} {\dot\varphi}_c
- M
\right)
\ \ ,
\label{CMvar-action}
\end{equation}
and the chart is thus canonical. The Hamiltonian $M$ in the new chart
is the unique solution in the interval $m_1+m_2 < M < \pi$ to
\begin{equation}
\frac{\left[ c_1^2 + c_2^2 - 2 c_1 c_2 \cos(M)
  \right]}{\sin^2(M)} = 
\left[
1 + \frac{p_{\varphi_c}^2}{{(1 - M/\pi)}^2 R_c^2}
\right]
\cosh^2 (P_c)
\ \ .
\label{ham-in-cm}
\end{equation}

By definition, $p_{\varphi_c}\ne0$ on~$\Gammaredplus$. We now extend
the chart $(R_c,\varphi_c;P_c,p_{\varphi_c})$ to $p_{\varphi_c}=0$ by
continuity, still maintaining the inequalities 
$R_c>0$ and~(\ref{cm-chart-ineq}). The action is given
by~(\ref{CMvar-action}), where $M$ is now the unique solution to
(\ref{ham-in-cm}) in the interval $m_1+m_2 \le M < \pi$; the lower
limit of this interval is achieved 
when $P_c = 0 = p_{\varphi_c}$. 
We denote the resulting extended reduced phase space by~$\Gammared$. 
The action (\ref{CMvar-action})
on $\Gammared$ correctly
reproduces all the classical solutions, including those with $S=0$: 
the spacetimes with colliding particles arise with 
$P_c
\ne 0 = p_{\varphi_c}$, and the static spacetimes arise with 
$P_c =0 = p_{\varphi_c}$. 

$\Gammared$~has dimension four.  In comparison, this 
is the dimension of the
phase space of the 
two-dimensional Newtonian two-body problem in the potential
$V(|\vec x_1 - \vec x_2|)$, after reduction 
to the centre-of mass frame.  It is further the dimension of a system of
two (say) free massive point particles in $(2+1)$-dimensional
Minkowski spacetime, after reduction to the centre-of-mass frame.
As discussed in section~\ref{sec:intro}, our anchoring the gravitating
system to the infinity is thus analogous to a reduction to the
centre-of-mass frame in Newtonian or special-relativistic physics.

The pair $(R_c,\varphi_c)$ provides a gravitational analogue of
the reduced position vector of the Newtonian two-body problem, and the 
conjugates $(P_c,p_{\varphi_c})$ provide a gravitational
analogue of the Newtonian reduced momentum. One aspect of this
analogue is the
recovery of the static solutions for $P_c =0 = p_{\varphi_c}$ and the 
solutions with colliding particles for $P_c \ne 0 = p_{\varphi_c}$. 
Another aspect is that in the spinning solutions,
recovered with $p_{\varphi_c}\ne0$, the particles are at their
smallest spatial separation when the `radial momentum' $P_c$ vanishes,
as seen from (\ref{Rc-def}) and~(\ref{transf-redmom-Pc}).

Because of the inequality~(\ref{cm-chart-ineq}), $\Gammared$~is a
genuine open subset of topology $\BbbR^3\times S^1$ of the cotangent
bundle over $\BbbR_+\times S^1 =
\{(R_c,\varphi_c)\}$. Qualitatively, (\ref{cm-chart-ineq}) says that
the momenta are bounded from above, and when 
(\ref{cm-chart-ineq}) approaches saturation, $M$~approaches its upper
bound~$\pi$. Discussion on this upper bound 
for more general matter sources 
can be found in \cite{ash-vara}. 

For further insight into the chart
$(R_c,\varphi_c;P_c,p_{\varphi_c})$, 
we consider three different limits. 

First, consider 
the slow motion limit. 
Expanding $M$ to quadratic order in $P_c$ and $p_{\varphi_c}$ yields 
\begin{mathletters}
\label{delta-nonrel}
\begin{equation}
M = m_1 + m_2 + 
\frac{1}{2m}
\left(P_c^2 + 
\frac{p_{\varphi_c}^2}{{[1 - (m_1+m_2)/\pi]}^2 R_c^2}
\right)
\ \ ,
\end{equation}
where 
\begin{equation}
m := 
{[\cot(m_1) + \cot(m_2)]}^{-1}
\ \ .
\end{equation}
\end{mathletters}%
Apart from the additive constant~$m_1+m_2$, 
(\ref{delta-nonrel}) is the Hamiltonian of a
nonrelativistic particle with mass $m$ on a cone with defect angle
$\delta_1+\delta_2 = 2(m_1 + m_2)$. 
$m$~is thus an ``effective mass'' 
that takes into account the quasistatic gravitational effects. When
$m_1$ and $m_2$ are both small, $m$~becomes the usual 
reduced mass for a free Newtonian two-particle system with the 
individual masses~$m_i$. We thus correctly recover in this limit the
free Newtonian two-body system in the centre-of-mass frame. 

Second, consider the limit in which the mass of particle $1$ is small 
but neither 
particle is moving close to the speed of light. 
To incorporate this, we assume that $P_c$ and $p_{\varphi_c}$ are
proportional to $m_1$ and expand $M$ to linear order
in~$m_1$, with the result 
\begin{equation}
M = m_2 + 
\sqrt{
m_1^2 + P_c^2 + 
\frac{p_{\varphi_c}^2}{{(1 - m_2/\pi)}^2 R_c^2}
}
\ \ .
\label{testparticle-ham}
\end{equation}
Apart from the additive constant~$m_2$, 
the expression (\ref{testparticle-ham})
is the familiar square-root Hamiltonian of a relativistic test particle 
with mass $m_1$ on the cone generated by particle~$2$ 
\cite{gibb-ruiz-vach}. We thus correctly recover 
the relativistic 
test particle limit for small~$m_1$. 
Further expanding (\ref{testparticle-ham}) to 
quadratic order in $P_c$
and~$p_{\varphi_c}$, with fixed 
$m_1$ and~$m_2$, 
yields the Hamiltonian of a nonrelativistic particle of mass $m_1$ on
a cone with defect angle~$2m_2$, 
in agreement with the limit of 
(\ref{delta-nonrel}) at small~$m_1$. 

Third, consider the limit in which 
the masses of both particles are small 
but neither 
particle is moving close to the speed of light. 
To incorporate this, we take $m_1$, $m_2$, 
$P_c$ and $p_{\varphi_c}$ all proportional to a small expansion
parameter 
and expand 
$M$ to linear order
in this parameter. The result is 
\begin{equation}
M = 
\sqrt{
m_1^2 + P_c^2 + 
\frac{p_{\varphi_c}^2}{R_c^2}
}
+ 
\sqrt{
m_2^2 + P_c^2 + 
\frac{p_{\varphi_c}^2}{R_c^2}
}
\ \ , 
\label{nongravpair-ham}
\end{equation}
which is the Hamiltonian of a special-relativistic test particle pair
in the centre-of-mass frame
\cite{loumatsch2}. 
Further expanding (\ref{nongravpair-ham}) to 
quadratic order in $P_c$
and~$p_{\varphi_c}$, with fixed 
$m_1$ and~$m_2$, 
yields the Hamiltonian of the 
free Newtonian two-body system in the centre-of-mass frame, 
in agreement with the limit of 
(\ref{delta-nonrel}) at small masses.

\section{Concluding remarks}
\label{sec:discussion}

In this paper we have anchored the Hamiltonian dynamics of a pair of
massive spinless point particles in $(2+1)$-dimensional Einstein
gravity to a 
conical spacelike infinity. This infinity is isometric
to that generated by a single massive but possibly spinning
particle, and assuming such an infinity to exist guarantees that the
spacetime is causally well behaved. 
We first described the two-particle spacetimes 
by relating the particle
trajectories to the asymptotic structure at the infinity. 
We then performed a 
Hamiltonian reduction of the first-order gravitational action under
boundary conditions 
adapted to this asymptotic structure. We found that the
reduced phase space $\Gammared$ is four-dimensional, and anchoring 
the dynamics to the conical infinity was seen to be analogous to
working in the centre-of-mass frame in Newtonian or flat spacetime
physics. In particular, we found on $\Gammared$ a canonical chart
in which the two configuration variables are analogous to the reduced
position vector of a Newtonian two-body system in the centre-of-mass
frame. 

In the Hamiltonian reduction, we took advantage of the
explicitly-known 
classical solutions and worked in variables that are
closely related to the constants of motion. 
We assumed in the reduction that the spacetime has
nonvanishing spin, and the resulting 
reduced phase space $\Gammaredplus$ 
thus only reproduced the spinning spacetimes. 
We then introduced 
on $\Gammaredplus$ a new canonical chart 
that is more closely related to
the configuration of the particles at 
a single moment of time, and only in this new chart did we
extend the reduced Hamiltonian system by continuity into the larger 
reduced phase space~$\Gammared$, in which also the nonspinning
spacetimes are correctly reproduced. 
While it seems likely that our reduction method could be directly 
extended to 
include the static spacetimes, 
the situation with the 
colliding-particle spacetimes is less clear, as the 
dynamics becomes indeterminate at the collisions. 
However, as the evolution of any point in our $\Gammared$ is
well defined for some finite interval of time, it seems
likely that the reduction to all of $\Gammared$ could be justified 
directly by
methods that are more tailored to initial data and less
reliant on the constants of motion. A~reduction of this type with a
second-order gravitational action 
has been recently discussed in 
\cite{meno-sem}. 

Although our Hamiltonian on $\Gammared$ was amenable to a classical
analysis, its functional form in the chart
$(R_c,\varphi_c;P_c,p_{\varphi_c})$ is determined only implicitly as
the solution to the transcendental equation~(\ref{ham-in-cm}).
Quantising the reduced Hamiltonian theory in these variables seems
thus to present a substantial 
challenge. A~more promising approach to quantisation
might open through reduction methods that are better adapted to initial
data and proceed step-by-step with partial gauge fixings, paying at
each step attention to the gauge symmetries still present in the
action and maintaining a freedom to choose gauges and variables that
yield simple charts on the partially reduced phase spaces.  Work in
this direction is in progress
\cite{loumatsch2}. 

Generalising the present work to more than two particles would appear
conceptually simple, although one may anticipate the complexity of the
reduced phase space to increase considerably with the number of
particles. Another generalisation would be to consider lightlike
particles
\cite{des-steif,des-mcc-steif}. 
Yet another direction would be to include a cosmological
constant and change the boundary conditions accordingly
\cite{steif-btz,matsch-creation,holst-matsch}, perhaps as motivated by
the CFT-AdS correspondence in string
theory~\cite{horo-itz,dani-etal,bala-ross}; in lineal gravity, an
analogous generalisation to a cosmological constant has been carried
out in \cite{Mann-lineal-coll}. We leave these issues subject to
future work.

\acknowledgments
We thank 
Ingemar Bengtsson, 
Nico Giulini 
and 
S\"oren Holst 
for helpful discussions, 
and two anonymous referees for 
helpful presentational suggestions and
bringing related work to our attention. 
For hospitality, both authors thank the University of
Stockholm, J.~L. thanks the University of Utrecht, and H.-J.~M. thanks 
the Max-Planck-Institut f\"ur Gravitations\-physik.

\newpage

\newpage

\newpage

\newpage

\newpage

\newpage

\newpage

\newpage

\newpage

\newpage

\newpage

\newpage

\begin{references}

\bibitem{carlip-book}
Carlip~S 
1998
{\it Quantum Gravity in 2+1 Dimensions} 
(Cambridge University Press, Cambridge, England)

\bibitem{deser1}
Deser~S, 
Jackiw~R
and
't~Hooft~G
1984
{\it Ann.\ Phys.\ \rm (N.Y.)} {\bf 152} 220 

\bibitem{deser2}
Deser~S 
and 
Jackiw~R
1984
{\it Ann.\ Phys.\ \rm (N.Y.)} {\bf 153} 405

\bibitem{thooft-cmp88}
't~Hooft~G
1988
{\it Commun.\ Math.\ Phys.\ 
\bf 117} 685

\bibitem{gott1}
Gott J~R  
1991
{\it Phys.\ Rev.\ Lett.\ 
\bf 66} 1126

\bibitem{ori}
Ori~A
1991
{\it Phys.\ Rev.\ \rm D 
\bf 44} R2214

\bibitem{deser-jack-th-prl}
Deser~S, 
Jackiw~R 
and
't~Hooft~G 
1992
{\it Phys.\ Rev.\ Lett.\ 
\bf 68} 267

\bibitem{carroll1}
Carroll S~M, 
Farhi~E, 
Guth~A 
and
Olum K~D 
1994
{\it Phys.\ Rev.\ \rm D 
\bf 50} 6190
[gr-qc/9404065]

\bibitem{carroll2}
Carroll S~M, 
Farhi~E
and 
Guth~A 
1992
{\it Phys.\ Rev.\ Lett.\ 
\bf 68} 263; 
Erratum, 
3368

\bibitem{cutler}
Cutler~C 
1992
{\it Phys.\ Rev.\ \rm D 
\bf 45} 487

\bibitem{thooft-closed1}
't~Hooft~G
1992
{\it Class.\ Quantum Grav.\ 
\bf 9} 1335

\bibitem{thooft-closed2}
't~Hooft~G
1993a
{\it Class.\ Quantum Grav.\ \bf 10} 1023

\bibitem{head-gott}
Headrick M~P 
and 
Gott J~R 
1994
{\it Phys.\ Rev.\ \rm D  \bf 50} 7244

\bibitem{carlip-scat}
Carlip~S 
1989
{\it Nucl.\ Phys.\ \rm B \bf 324} 106 

\bibitem{cap-cia-val-plb}
Cappelli~A, 
Ciafaloni~M
and 
Valtancoli~P
1991
{\it Phys.\ Lett.\ \bf 273B} 431
[hep-th/9110020]

\bibitem{cap-cia-val-npb}
Cappelli~A, 
Ciafaloni~M
and 
Valtancoli~P
1992
{\it Nucl.\ Phys.\ \rm B \bf 369} 669

\bibitem{thooft-quantum}
't~Hooft~G
1993b
{\it Class.\ Quantum Grav.\ \bf 10} 1653
[gr-qc/9305008]

\bibitem{bel-cia-val-plb}
Bellini~A, 
Ciafaloni~M
and 
Valtancoli~P
1995a
{\it Phys.\ Lett.\ \bf 357B} 532
[hep-th/9507076] 

\bibitem{bel-cia-val-npb95}
Bellini~A, 
Ciafaloni~M
and 
Valtancoli~P
1995b
{\it Nucl.\ Phys.\ \rm B \bf 454} 449
[hep-th/9507077] 

\bibitem{bel-cia-val-npb96}
Bellini~A, 
Ciafaloni~M
and 
Valtancoli~P
1996
{\it Nucl.\ Phys.\ \rm B \bf 462} 453
[hep-th/9511207]

\bibitem{welling-noncomm}
Welling~M
1997 
{\it Class.\ Quantum Grav.\ \bf 14} 3313
[gr-qc/9703058]

\bibitem{welling-winding}
Welling~M
1998
{\it Class.\ Quantum Grav.\ \bf 15} 613
[gr-qc/9704067] 

\bibitem{matsch-well}
Matschull H-J 
and 
Welling~M
1998
{\it Class.\ Quantum Grav.\ \bf 15} 2981
[gr-qc/9708054]

\bibitem{meno-sem}
Menotti~P 
and 
Seminara~D 
1999
``ADM approach to 2+1 dimensional gravity coupled to particles'' 
Preprint IFUP-TH~39/99, LPTENS-99/24, hep-th/9907111. 

\bibitem{MTW-mass-angmom}
Misner C~W, 
Thorne K~S, 
and 
Wheeler J~A
1973
{\it Gravitation\/} (Freeman, San Francisco) 
Chapter 19

\bibitem{regge-teitel}
Regge~T 
and 
Teitelboim~C 
1974
{\it Ann.\ Phys.\ \rm (N.Y.) \bf 88} 286

\bibitem{beig-om}
Beig~R 
and 
\'{o}~Murchadha~N
1987
{\it Ann.\ Phys.\ \rm (N.Y.) \bf 174} 463

\bibitem{henneaux-conical}
Henneaux~M
1984
{\it Phys.\ Rev.\ \rm D \bf 29} 2766

\bibitem{ash-vara}
Ashtekar~A 
and 
Varadarajan~M
1994
{\it Phys.\ Rev.\ \rm D \bf 50} 4944
[gr-qc/9406040]

\bibitem{achu}
Ach\'ucarro~A 
and 
Townsend P~K
1986
{\it Phys.\ Lett.\ \bf 180B} 85

\bibitem{witten1}
Witten E
1988
{\it Nucl.\ Phys.\  \rm B \bf 311} 46

\bibitem{AAbook2}
Ashtekar~A
1991
{\it Lectures on Non-Perturbative Canonical Gravity\/}
(World Scientific, Singapore) 
Chapter 17

\bibitem{romano}
Romano J~D
1993
{\it Gen.\ Rel.\ Grav.\ \bf 25} 759
[gr-qc/9303032]

\bibitem{loumar-wittentorus}
Louko~J 
and 
Marolf D~M 
1994
{\it Class.\ Quantum Grav.\ \bf 11} 311 
[gr-qc/9308018]



\bibitem{gibb-ruiz-vach}
Gibbons G~W, 
Ruiz Ruiz F 
and 
Vachaspati~T 
1990
{\it Commun.\ Math.\ Phys.\  \bf 127} 295 

\bibitem{loumatsch2}
Louko~J
and 
Matschull H-J 
in preparation

\bibitem{des-steif}
Deser~S
and
Steif A~R
1992
{\it Class.\ Quantum Grav.\ \bf 9} L153
[hep-th/9208018]

\bibitem{des-mcc-steif} 
Deser~S, 
McCarthy~J 
and
Steif A~R
1994
{\it Nucl.\ Phys.\ \rm B \bf 412} 305
[hep-th/9307092]

\bibitem{steif-btz}
Steif A~R 
1996
{\it Phys.\ Rev.\  \rm D  \bf 53} 5527 
[gr-qc/9511053]

\bibitem{matsch-creation}
Matschull H-J 
1999
{\it Class.\ Quantum Grav.\ \bf 16} 1069
[gr-qc/9809087]

\bibitem{holst-matsch}
Holst~S 
and 
Matschull H-J 
1999
{\it Class.\ Quantum Grav.\ \bf 16} 3095
[gr-qc/9905030]

\bibitem{horo-itz}
Horowitz G~T 
and
Itzhaki~N 
1999
{\it JHEP\/ \bf 9902} 010
[hep-th/9901012]

\bibitem{dani-etal}
Danielsson U~H, 
Keski-Vakkuri~E
and 
Kruczenski~M 
1999
{\it Nucl.\ Phys.\ \rm B \bf 563} 279
[hep-th/9905227]

\bibitem{bala-ross}
Balasubramanian~V 
and 
Ross S~F 
2000
{\it Phys.\ Rev.\  \rm D  \bf 61} 044007
[hep-th/9906226]

\bibitem{Mann-lineal-coll}
Ohta~T
and
Mann R~B 
1996
{\it Class.\ Quantum Grav.\ \bf 13}
2585
[gr-qc/9605004]; 
Mann R~B 
and 
Ohta~T
1997a
{\it Phys.\ Rev.\ \rm D  \bf 57} 4723
[gr-qc/9611008]; 
Mann R~B 
and 
Ohta~T
1997b
{\it Class.\ Quantum Grav.\ \bf 14} 1259 
[gr-qc/9607016]; 
Mann R~B, 
Robbins~D
and 
Ohta~T
1999a
{\it Phys.\ Rev.\ Lett.\ \bf 82} 3738
[gr-qc/9811061]; 
Mann R~B, 
Robbins~D
and 
Ohta~T
1999b
{\it Phys.\ Rev.\ \rm D  \bf 60} 104048
[gr-qc/9906066] 




\end{references}
\end{document}